\documentclass[sigconf]{acmart}
\usepackage[ruled,linesnumbered]{algorithm2e}
\usepackage{multirow}
\usepackage{graphicx}
\usepackage[normalem]{ulem}
\useunder{\uline}{\ul}{}
\usepackage{caption}
\usepackage{subfig}
\usepackage{enumitem}
\AtBeginDocument{%
  }

\renewcommand\footnotetextcopyrightpermission[1]{}

\acmConference[]{}{}{}
\acmPrice{15.00}
\acmISBN{978-1-4503-XXXX-X/18/06}

\begin{document}

\title{BotSCL: Heterophily-aware Social Bot Detection with Supervised Contrastive Learning}

\author{Qi Wu$^1$, Yingguang Yang$^1$, Buyun He$^1$, Hao Liu$^1$, Renyu Yang$^2$ and Yong Liao$^1$ }
\affiliation{%
  \institution{$^1$University of Science and Technology Of China; $^2$Beihang University.}
  \country{China}
}
\email{qiwu4512@mail.ustc.edu.cn}


\renewcommand{\shortauthors}{Wu and Yang, et al.}

\begin{abstract}
    Detecting ever-evolving social bots has become increasingly challenging. Advanced bots tend to interact more with humans as a camouflage to evade detection. While graph-based detection methods can exploit various relations in social networks to model node behaviors, the aggregated information from neighbors largely ignore the inherent heterophily, i.e., the connections between different classes of accounts. Message passing mechanism on heterophilic edges can lead to feature mixture between bots and normal users, resulting in more false negatives. In this paper, we present BotSCL, a heterophily-aware contrastive learning framework that can adaptively differentiate neighbor representations of heterophilic relations while assimilating the representations of homophilic neighbors.  Specifically, we employ two graph augmentation methods to generate different graph views and design a channel-wise and attention-free encoder to overcome the limitation of neighbor information summing. Supervised contrastive learning is used to guide the encoder to aggregate class-specific information. Extensive experiments on two social bot detection benchmarks demonstrate that BotSCL outperforms baseline approaches including the state-of-the-art bot detection approaches, partially heterophilic GNNs and self-supervised contrast learning methods.
\end{abstract}

\ccsdesc[500]{Computing methodologies~Machine learning}
\ccsdesc[500]{Security and privacy~Social network security and privacy}

\keywords{homophily and heterophily, Social bot detection, supervised contrastive learning, graph neural networks}


\maketitle

\section{Introduction}
Social bots are automated accounts that are often used for malicious purposes, such as spreading misinformation \cite{cresci2020decade}, promoting extremism \cite{hamdi2022mining}, and electoral interference \cite{deb2019perils,ferrara2017disinformation}. Bots have been widely existing in social networks and continuously evolve to tackle emerging detection techniques. A variety of advanced bot detection technologies safeguard the environment of social networks. Approaches based on extracting distinctive characteristics from Twitter accounts typically extracted tweets \cite{kudugunta2018deep}, metadata \cite{beskow2019its,yang2020scalable}, and temporal features \cite{chavoshi2017temporal} and fed into various classifiers. Deep neural networks with different architectures are further designed to improve classification performance. However, they fail to model the diverse relationships (e.g., following, commenting, etc.) between social accounts. Recent advancements in graph neural networks based approaches \cite{ali2019detect,feng2021botrgcn,yang2023fedack,feng2022heterogeneity,yang2022rosgas} can better capture semantic relationship information. A common practice is constructing a heterogeneous graph that contains different relations before using a relational graph transformer to aggregate both intra-relational and inter-relational information. Account feature information and topological information can be therefore co-utilized to obtain richer semantic embeddings with more comprehensive information.

\begin{figure}[t]
    \centering
    \includegraphics[width=0.9\linewidth]{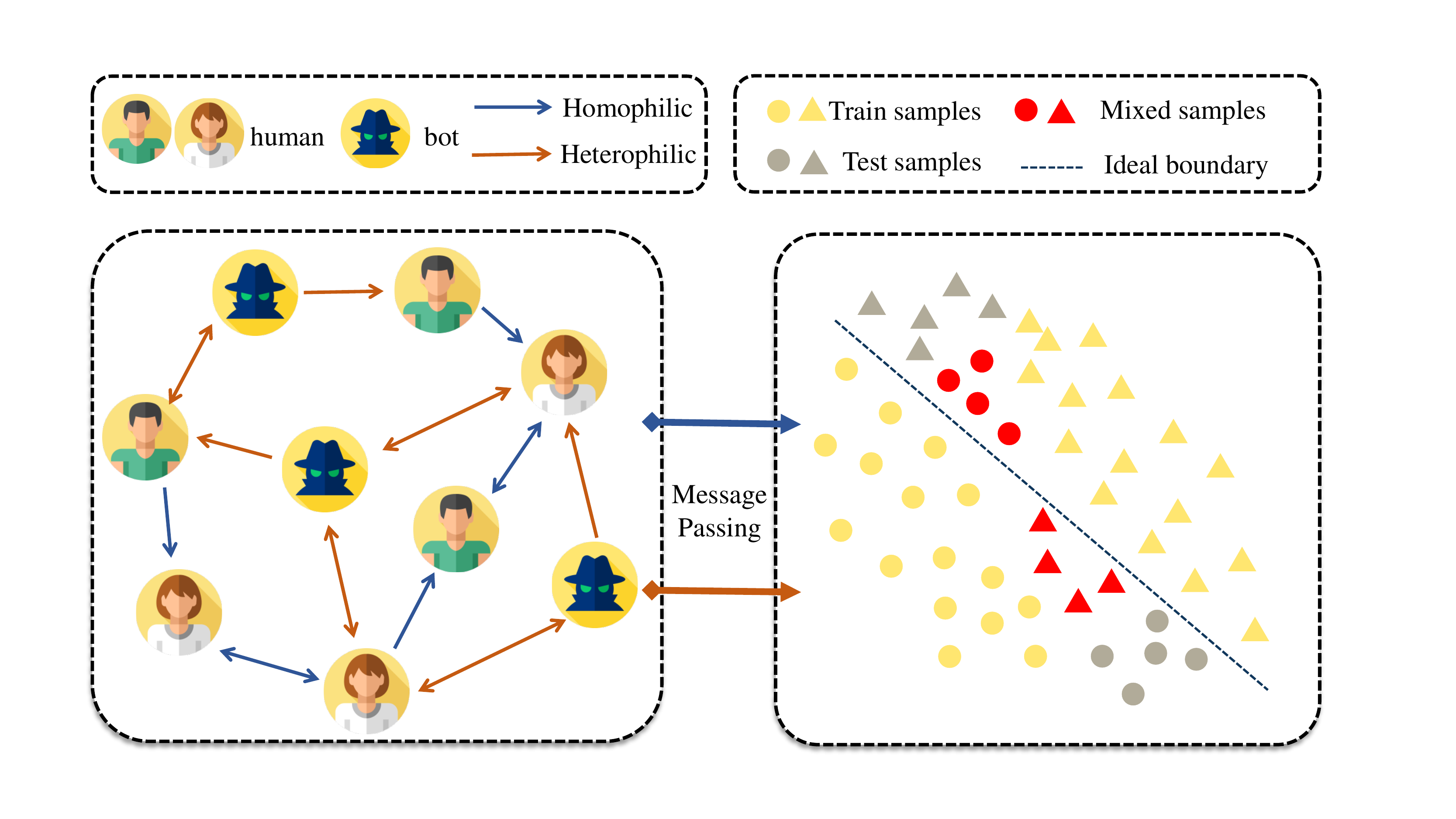}
    \caption{Illustration of Heterophily Influence. Message passing on heterophilic edges leads to feature mixing and classification boundary shift in social bot detection.}
    \Description{On the left side is an illustration of a graph that includes both homophilic and heterophilic relations and on the right side is the distribution of node features after message passing.}
    \label{fig:intro}
\end{figure}

However, most of the existing graph-based methods overlook the negative impact of heterophily (i.e., the connections between different classes of accounts). Recent investigations \cite{williams2020homophily,des2022detecting} have revealed that bots intentionally interacting more with humans inherently lead to adversarial properties of social bots: increased influence of particular information and intentional evasion of detection \cite{le2022socialbots}. Social bots can easily establish heterophilic relationship with human beings by following normal user accounts or replying to tweets of normal users. The information aggregation mechanism used in the existing graph-based detection methods -- often seen as the sum of node representations in the neighborhood -- tends to make neighboring node representations similar. As shown in Figure \ref{fig:intro}, once homophilic and heterophilic interactions co-exist, message passing through heterophilic edges will assimilate node representations to those of the opposite class, leading to less distinguishable representations. As a result, such mixed representations trained with their original labels ends up with shifted classification boundary and increasing false positives. 

Therefore, effective detection should make node representation pertaining to the same class as similar as possible, whilst staying away from other classes. Most of the existing works ~\cite{ali2019detect,feng2021botrgcn,feng2022heterogeneity} can only assimilate the representations of nodes belonging to different categories during message passing and thus harmful for classification ~\cite{shi2022h2}. That is to say, we first need to further extend previous relational graph neural networks in a way that enables them to freely and adaptively promote both intra-class similarity and inter-class differentiation among neighboring nodes. Supervised contrastive learning \cite{khosla2020supervised} is intrinsically the effective means to characterizes inter-class discrimination. With this optimization objective, non-adjacent nodes of the same class are considered positive pairs, and nodes tend to aggregate global class-specific information rather than local information. In this way, it ensures that the node representations after message passing are closer to the category centers which is beneficial for weight matrix to classify.


In this paper, we propose BotSCL, a supervised contrastive social bot detection framework that co-considers homophilic and heterophilic relations. We devise two graph augmentation methods -- including  feature augmentation and topological structure augmentation -- to obtain different graph views. We then propose an encoder to aggregate similar and different information on a per feature channel basis. Supervised contrastive learning is exploited in a cross-view manner to obtain class-consistent representations between different graph views. Consequently, it compels the encoder to assimilate representations of homophilic neighbors while differentiating representations from heterophilic neighbors. 

In particular, this paper makes the following contributions:
\begin{itemize}
    \item We first introduce and reveal the negative impact of heterophily on social bot detection, and experimentally validate it.
    \item We propose a detection framework that aggregates similar and differentiable information with the guidance of supervised contrastive learning.
    \item We conduct experiments on two social bot detection benchmark datasets. The results show that our model consistently outperforms previous state-of-the-art methods.
\end{itemize}

\section{Related Work}
In this section, we will discuss relevant research on graph-based social bot detection, graph neural networks for heterophilic graphs, and contrastive learning (e.g., \cite{feng2021satar,feng2022heterogeneity,moghaddam2022friendship}).
\subsection{Graph-based Social Bot Detection} 

Previous approaches to social bot detection primarily involve manual analysis of collected data and extraction of distinctive characteristics for input into diverse classifiers\cite{yang2022botometer}. Subsequently, deep neural networks with diverse architectures are developed to enhance classification performance\cite{kudugunta2018deep, wu2021novel}. However, the detectability of these characteristics is vulnerable to imitation and evasion by social bots, rendering them ineffective over time. To tackle the challenge of bot disguise, graph-based social bot detection methods have been extensively studied and shown great success in social bot detection with the advent of benchmark datasets that incorporate graph information\cite{feng2021twibot, feng2022twibot}.

\cite{ali2019detect} is the first attempt to introduce graph convolutional networks to take advantage of both the features of the accounts and the structure of the relation graph, which takes Twitter users as nodes. Satar \cite{feng2021satar} utilizes graph convolutional networks in a feature engineering manner and employs self-supervision to detect social robots. Relational graph convolutional networks (R-GCNs) \cite{schlichtkrull2018modeling} were used by \cite{feng2021botrgcn} to aggregate information from different relations and \cite{feng2022heterogeneity} later improved it with additional relations and apply graph transformer to better aggregate information from neighbors adaptively. \cite{yang2022rosgas} proposed RoSGAS, a framework that leverages heterogeneous information network  to effectively model multiple entities and relationships within a social network and performs subgraph embedding with reinforcement learning for social bot detection.

All of these methods are based on the assumption that both humans and bots tend to interact more with the same class, and classification benefits from smoothing the representations of neighboring nodes. However, in reality, advanced bots can easily escape graph-based detection by actively interacting with humans because none of these approaches take heterophily into account. In this paper, we propose a bot detection framework that recognizes the negative impact of heterophilic relations between bots and humans for more effective user representations to enhance the performance of bot detection.

\subsection{Heterophilic GNNs} 
Due to the widespread existence of heterophily, graph neural networks for graphs with heterophily have received significant attention in recent years. Essentially, there are mainly two kinds of approach: 1) Aggregation of non-local neighbor information. For example, some techniques gather information from higher-order neighbors \cite{abu2019mixhop,zhu2020beyond} and potential neighbors of the same class \cite{pei2020geom,wang2022powerful} to obtain more intra-class information. 2) Adaptive Message Passing. FAGCN \cite{bo2021beyond} simultaneously aggregates high- and low-frequency information. GPRGNN \cite{chien2020adaptive} employs learnable weights for information aggregated from distinct hop neighbors. \cite{luan2021heterophily} argues that not all heterophily edges are harmful for classification and proposes Adaptive Channel Mix (ACM) to adaptively aggregate self-information, low-frequency information, and high-frequency information. Despite the emergence of numerous heterophilic GNNs, these methods primarily focus on single-relation and undirected simple graphs, particularly spectral-based GNNs. In social bot detection, bots exhibit a higher tendency towards heterophily, while humans display a higher inclination towards homophily, respectively. Therefore, directly applying heterophilic GNNs to social bot detection may not yield optimal results. 

\subsection{Contrastive Learning} 
Contrastive learning aims to learn an encoder that can generate representations consistent with different views by attracting positive pairs and repelling negative pairs. Graph contrastive learning (GCL) extends the technique to graph domain and learns the representations of nodes in a self-supervised manner due to rich information implicitly existing in the connections between nodes. GRACE \cite{zhu2020deep} applies edge removal and feature mask to generate augmented views and treat the same node in different views as a positive pair. DGI \cite{velickovic2019deep} learns node representations by maximizing the mutual information between local and global embeddings. However, self-supervised contrastive learning methods face class collision problem --  the representations of similar samples may be far apart, while the representations of dissimilar samples may be close \cite{zheng2021weakly}. Supervised contrastive learning \cite{khosla2020supervised} firstly applied in computer vision field treats intra-class images as positive pairs, while inter-class images as negative pairs. Thus, embeddings from the same class are pulled closer than embeddings from different classes. Because this property is opposite to the feature mixing caused by message propagation on heterophilic edges, we utilize supervised contrastive loss to train an encoder that can adapt to both homophilic and heterophilic edges simultaneously to aggregate information beneficial for further classification.

\section{PRELIMINARIES}

We first formulate the task of graph-based social bot detection, and present an approach to measure the homophily and heterophily degree. To aid discussion, Table \ref{tab:notation} depicts the notations used in the paper.
\vspace{0.5\baselineskip}

\noindent \textbf{Definition 3.1. Graph-based Social Bot Detection.} Previous graph-based approaches for detecting social bots consider social networking platform (e.g., Twitter) accounts as nodes and interactive behaviors such as follower and following as edges. Social bot detection can thus be considered as semi-supervised node classification on an attributed multi-relational graph. We define this graph as $\mathcal{G}=\left\{\mathcal{V},\mathcal{E}, \mathcal{X}\right\}$, where $\mathcal{V}=\{v_{i} \mid i=1,2, \ldots, n\}$ is the set of all nodes, $\mathcal{E}$ represents the set of edges $\mathcal{E}=\{\mathcal{E}_{r} = r\in\ 1,2, \ldots, R\}$ formed by different relations and $\mathcal{X}$ is the feature matrix, each row of which represents the feature vector of the corresponding node. Total detection process is to use the graph $\mathcal{G}$ and the labels of training nodes $\mathrm{Y}_{\text {train }}$ to predict the labels of test nodes $\hat{\mathrm{Y}}_{\text {test }}$:
\begin{equation}
    f\left(\mathcal{G}, \mathrm{Y}_{\text {train }}\right) \rightarrow \hat{\mathrm{Y}}_{\text {test }}.
\end{equation}

\vspace{0.5\baselineskip}
\noindent \textbf{Definition 3.2. Homophily and Heterophily Measure.} To gain a deeper insight into the extent of homophily and heterophily in social bot detection, we further consider different relations, directionality, and classes on top of the previous metric \cite{zhu2020beyond}. The class-aware homophily and heterophliy ratio of given graph $\mathcal{G}$ in terms of relation $r$ can be defined as:
\begin{equation}
\begin{aligned}
    & homo(\mathcal{G}, r , c)=\frac{\left|e_{ij} \in \mathcal{E}_{r}: y_{i}=c, y_{j}=c\right|}{\left|e_{ij} \in \mathcal{E}_{r}: y_{i}=c\right|}, \\
    & hetero(\mathcal{G}, r , c) = 1- homo(\mathcal{G}, r , c),\\ 
\end{aligned}
\end{equation}
where $e_{ij}$ represents the directed edge from node $v_i$ to $v_j$ and $c_{i}$ is the label of node $v_i$, 0 for human and 1 for bot. In this measurement setup, the edges are considered as the active behaviors of their start nodes. For the reason that benchmark datasets may only include a small proportion of labeled nodes, we only consider edges where both the starting and ending nodes have labels in our calculations.

\begin{table}[t]
\centering
\caption{Glossary Of Notations} 
\label{tab:notation}
\begin{tabular}{@{}r|l@{}}
\toprule[0.5pt]
\toprule[0.5pt]
\textbf{Notation}         & \textbf{Description} \\ 
\midrule[0.5pt]
$\mathcal{G}$; $\mathcal{V}$; $\mathcal{E}$; $\mathcal{X}$ & Graph; Node set; Edge set;  Node feature matrix\\
$v$; $e$; & The node $v$; The edge $e$;  \\
$e_{ij}$; $x_{i}^{t}$   & The edge between node $v_i$ and $v_j$; Node $v_i$'s feature  \\
$r$; $R$   & The relation $r$; Total number of relations \\
$\mathcal{E}_{r}$ & Set of edges formed by the relation $r$ \\
$h_{i}^{\{l\}}$     & Hidden state of node $v_i$ at layer $l$ \\
$y_{i}$; $\mathrm{Y}$      & Label of node $v_i$; Node label set\\
$\hat{y}_{i}$; $\hat{\mathrm{Y}}$      & Predicted label for node $v_i$; Predicted label set\\
$c$   & A class of nodes in social network graph\\
$homo(\cdot)$         & A function to compute homophily ratio\\
$hetero(\cdot)$         & A function to compute heterophily ratio\\
$\mathcal{C(\cdot)}$      & Class-aware graph augmentation function\\
$m_{ij}^{\mathcal{E}}$; $M^{\mathcal{E}}$             & The mask value of $e_{ij}$; Mask value of $\mathcal{E}$ \\
$\mathcal{B}(\cdot)$      & Bernoulli distribution \\
$W_r^{\{l\}}, b^{\{l\}}$      & Learnable model parameters at layer $\{l\}$  \\
$Q^{\{l\}}$; $K^{\{l\}}$      & Query weight for layer $\{l\}$ ; Key weight for layer $\{l\}$ \\
$\mathcal{N}_{r}(v_i)$      & The neighborhood of node $v_i$ with regard to relation $r$ \\
$\alpha_{ij}^{\{l\}}$      & The attention co-efficient for $e_{ij}$ at layer $\{l\}$ \\
$\odot$      & Hadamard product operation \\
$\sigma$      & Nonlinear activation function \\
$z_i$      & The representation of node $v_i$ \\
$cos(\cdot,\cdot)$      & Cosine similarity function \\
$\tau$      & Temperature coefficient \\
\bottomrule[0.5pt]
\bottomrule[0.5pt]
\end{tabular}
\end{table}

\section{Methodology}
\begin{figure*}[t]
    \centering
    \includegraphics[width=\linewidth]{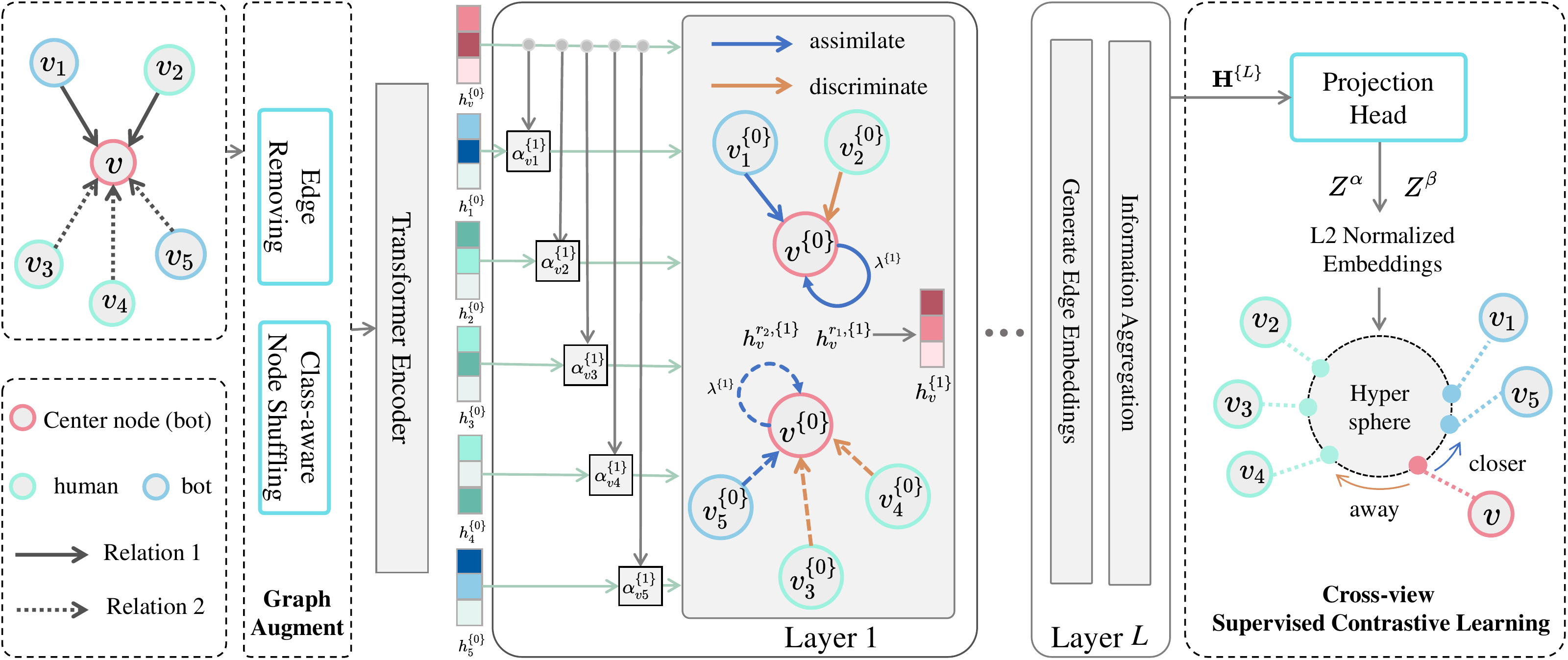}
    \caption{The proposed BotSCL framework.}
    \Description{The total pipeline of BotSCL.}
    \label{fig:framework}
\end{figure*}

This section presents the details of BotSCL. Same with the general framework of contrastive learning, our framework consists of a graph augmentor, an encoder, and a contrastive loss. The total pipeline of BotSCL is shown in Figure \ref{fig:framework}. First, we use two graph augmentation methods to generate two graph views. Then, nodes aggregate similar information from homophilic neighbors and adaptively discriminate representations from heterophilic neighbors for each relation. Lastly, node representations are learned through supervised contrastive learning in a cross view manner. Our goal is to train an encoder that can learn effectively from both heterophilic and homophilic edges.

\subsection{Augmentor}
Graph contrastive learning employs multiple graph augmentation strategies, including edge addition, feature masking, and personalized pagerank diffusion, to generate various graph views \cite{you2020graph}. It should be noted that not all graph augmentation methods can be universally applicable to graphs with heterophily. \cite{yang2023contrastive} uses a low-pass filter and high-pass filter to generate graph views for self-supervised contrastive learning. However, this approach may result in the absence of directional information on relations and higher-order neighbor information. \cite{liu2022beyond} first determines whether each edge belongs to within-class or between-class, and then generates a homophilic view and a heterophilic view for contrastive learning. However, error propagation will inevitably happen. This paper aims to use supervised contrastive learning loss to enable simultaneous perception of both heterophilic and heterophilic edges, instead of employing extra differentiation between the two.

As supervised contrastive learning leverages label information during the training process, we propose a class-aware node shuffling (CNS) graph augmentation method that randomly swaps nodes belonging to the same class. By using this approach, one can learn representations that are invariant to the neighborhood without excessively disrupting the graph structure. This graph augmentation method can be implemented through intra-class feature swapping:
\begin{equation}
    \widetilde{\mathcal{X}}=\mathcal{C}(\mathcal{X},\mathrm{Y}_{train}).
    \label{eq:aug1}
\end{equation}
Additionally, we also employed a traditional graph augmentation method, edge removing (ER), to augment the graph topology. The edge removal method can be formulated as follows:
\begin{equation}
    \tilde{\mathcal{E}}= \mathcal{E} \odot M^\mathcal{E},
    \label{eq:aug2}
\end{equation}
where $\mathcal{E}$ is total edge set, each element of $M^\mathcal{E}$, $m_{ij}^{\mathcal{E}}$ stands for the mask value of edge $e_{ij}$ and $m^{\mathcal{E}}_{ij}\sim\mathcal{B}(1-pe)$, $pe$ is the probability to remove. Through the aforementioned two graph augmentation methods, both the features and topological structure of the original graph are augmented.

\subsection{Encoder}
\subsubsection{Feature Fusion.}
Unlike other node classification tasks, the input for the social bot detection task consists of features of different types and dimensions. Therefore, it is necessary to use a multi-layer perceptron (MLP) for each type of feature to align the dimensions of feature vectors. Previous methods \cite{feng2021botrgcn, feng2022heterogeneity} further concatenate these different types of feature vectors and use another MLP to obtain the input of the graph convolutional layer. 

In this study, we use TransformerEncoder \cite{vaswani2017attention} for feature fusion by treating the feature vectors of different types as token embeddings:
\begin{equation}
    x_i^{0}= TransformerEncoder([x_i^{1},x_i^{2},\dots,x_i^{t}]),
    \label{eq:input1}
\end{equation}
where $x_i^{t}$ stands for the feature vector of $t$ type
of node $v_i$ and $x_i^{0}$ is the concatenation of the TransformerEnocder outputs. On top of that, we apply another MLP to $x_i^{0}$ and obtain the input of the graph convolutional layer $h_i^{\{0\}}$:
\begin{equation}
    h_i^{\{0\}} = \sigma (W_Ix_i^{0}+b_I),
    \label{eq:input2}
\end{equation}
where $W_I$, $b_I$ are learnable parameters and $\sigma(\cdot)$ is a nonlinear activation function. Thanks to the mechanisms of self-attention and residual connections in Transformer, we can obtain more semantically enriched node input representations.

\subsubsection{Information aggregation.}
After obtaining different views, BotSCL needs to use an encoder that can aggregate information from similar and dissimilar neighbors in a distinguishable manner to obtain node representations. In the spectral domain, GNNs based on the homophily assumption can be viewed as low-pass filters, while previous work \cite{luan2021heterophily} on graphs with heterophily has shown that high-pass filters that capture differential information are more effective for modeling heterophilic connections. On the other hand, low-pass filtering can be achieved by aggregating information from neighboring nodes, while high-pass filtering can be achieved by differentiating neighbor representations in the spatial domain. This can be formulated as: 
\begin{equation}
    \begin{aligned}
        \left(h_{i}^{\{l\}}\right)_{Low}=W^{\{l-1\}} \sum _{j \in \mathcal{N}_{(i)}}\left(h_{i}^{\{l-1\}}+h_{j}^{\{l-1\}}\right), \\
        \left(h_{i}^{\{l\}}\right)_{High}=W^{\{l-1\}} \sum _{j \in \mathcal{N}_{(i)}}\left(h_{i}^{\{l-1\}}-h_{j}^{\{l-1\}}\right).
    \end{aligned}
\end{equation}

Inspired by the above, we design a channel-wise self-attention mechanism to adaptively aggregate similar information from homophilic neighbors and differential information from heterophilic neighbors. Specifically, given a central node $v_i\in V$ and its arbitrary neighbor $v_j \in N_r(v_i)$, we first use a linear transformation and a separate element-wise multiplication across channels to obtain the query and key:
\begin{equation}
    \begin{aligned}
        q_i^{\{l\}}=W_A^{\{l\}}h_i^{\{l-1\}}\odot Q^{\{l\}}, \\
        k_j^{\{l\}}=W_A^{\{l\}}h_j^{\{l-1\}}\odot K^{\{l\}},
    \end{aligned}
    \label{eq:att1}
\end{equation}
where $W_A^{\{l\}}\in \mathbb{R}^{d_{l-1} \times d_{l-1}}$ is the weight martix of layer $l$, $Q^{\{l\}}, K^{\{l\}} \in \mathbb{R}^{1 \times d_{l-1}}$ are weight vectors for query and key and $\odot$ denotes the Hadamard product operation. $q_j^{\{l\}}$ and $k_i^{\{l\}}$ can also be calculated in the same way.

Then we calculate the channel-wise and pass-free attention coefficient $\alpha_{ij}^{\{l\}}$ for edge $e_{ij}$:
 \begin{equation}
     \alpha_{ij}^{\{l\}}=tanh(\frac{q_i^{\{l\}}\odot k_j^{\{l\}}+q_j^{\{l\}}\odot k_i^{\{l\}}}{2}).
     \label{eq:att2}
 \end{equation}
It is worth noting that the obtained $\alpha_{ij}^{\{l\}}$ using the above calculation method is direction- and relation-agnostic and can be also seen as the embedding of edge $e_{ij}$. Furthermore, due to the use of the tanh activation function, any element in $\alpha_{ij}^{\{l\}}$ is in the range of [-1, 1], which breaks the previous restriction on the sum of neighbor information.

Finally, we aggregate information from the neighbors using the generated channel-wise weights $\alpha_{ij}^{\{l\}}$ to obtain $l$ layer node representation $h_{i}^{\{l\}}$:
\begin{equation}
        h_{i}^{r, \{l\}}=W_{r}^{\{l\}}(\lambda^{\{l\}}h_{i}^{\{l-1\}}+\sum_{j\in N_{r}\left(i\right)}\frac{\alpha_{ij}^{\{l\}}}{| N_{r}(i) |}\odot h_{j}^{\{l-1\}}),
    \label{eq:att3}
\end{equation}
\begin{equation}
    h_{i}^{\{l\}}=\frac{1}{R}\sum_{r=1}^{R}h_{i}^{r, \{l\}},
    \label{eq:att4}
\end{equation}
where $W_{r}^{\{l\}} \in \mathbb{R}^{d_{l-1} \times d_{l}}$ is the weight matrix for relation $r$, and 
$| N_{r}(v_i) |$ is the number of neighbor nodes on relation $r$. Following \cite{bo2021beyond}, we apply a hyperparameter $\lambda^{\{l\}}$ to preserve the information of the node itself. RGT \cite{feng2022heterogeneity} uses an attention mechanism to fuse information from different relations, but here we trivially take the average of information from different relations to avoid information missing.

\subsection{Contrastive Loss}
Following the traditional contrastive learning framework, we use a projection head consisting of two MLP layers to obtain $z_i$:
\begin{equation}
    z_i=W_2\sigma(W_1h_i^{\{L\}}+b_1)+b_2,
    \label{eq:proj}
\end{equation}
where $h_i^{L}$ is the output of last layer $L$. Thus, we can obtain projections $z_i^{\alpha}$ and $z_i^{\beta}$ of node $v_i$ in two graph views $\mathcal{G}^\alpha$ and $\mathcal{G}^\beta$ in respect. 

Next, we use supervised contrastive learning as the objective loss function for training. However, due to the fact that supervised contrastive learning treats all nodes of the same class in different views as positive pairs, it is prone to overfitting, where the representations of all same-class training nodes become too similar and cannot generalize to test nodes. To avoid overfitting, we use supervised contrastive learning in a cross-view manner. For a set of N samples randomly sampled from the training nodes, we obtain the projection of each node through the graph augmentation, encoder, and projection head mentioned above. Taking node $v_i$ in graph view $\mathcal{G}^\alpha$ as an example, we treat its projection and the projections of same-class nodes in the other view as positive pairs, and those of different-class nodes as negative pairs, to calculate the contrastive loss between them:
\begin{equation}
    \mathcal{L}_i^\alpha=-\frac{1}{N_{y_i}}\sum_{j=1}^N1_{y_i=y_j}\cdot log\frac{e^{cos(z_i^\alpha,z_j^\beta)/\tau}}{\sum\limits_{k=1}^N e^{cos(z_i^\alpha, z_k^\beta)/\tau}}, 
    \label{eq:con1}
\end{equation}
where $N_{y_i}$ represents the number of samples in the same class as node $v_i$ among $N$ samples, $cos(\cdot , \cdot )$ function is used to calculate the cosine similarity, and $\tau$ is the temperature coefficient which can regulate the degree of distribution uniformity.

Finally, we calculate the loss for all nodes in the sampled set of two views in the same way, and take the average:
\begin{equation}
    \mathcal{L}=\frac{1}{2N}\sum_{i=1}^N(\mathcal{L}_i^\alpha +\mathcal{L}_i^\beta).
    \label{eq:con2}
\end{equation}

\begin{algorithm}[!b]
    \caption{The first training process of BotSCL}
    \label{algorithm}
    \SetKwInOut{Input}{Input}
    \SetKwInOut{Output}{Output}
    \SetAlgoLined
    \Input{a directed and multi-relation graph $\mathcal{G}=\left\{\mathcal{V},\mathcal{E}, \mathcal{X}\right\}$, the labels of train nodes $\mathrm{Y}_{\text{train}}$, Training epochs $N_{epochs}$, the number of layers in the encoder $L$}
    \Output{node representations $\mathbf{H}$}
    initialization\;
    generate two graph views $\mathcal{G}^\alpha$ and $\mathcal{G}^\beta$ $\leftarrow$ Equation (\ref{eq:aug1}-\ref{eq:aug2})\; 
    \For{$e=1,\cdots,N_{epochs}$}{
        obtain $\mathbf{H}^{\{0\}}$ $\leftarrow$ Equation (\ref{eq:input1}-\ref{eq:input2})\;
        \For{each graph view}{
            \For{$l=1,\cdots,L$}{
                $q_i^{\{l\}}, q_j^{\{l\}}, k_i^{\{l\}},k_i^{\{l\}} \leftarrow$ Equation (\ref{eq:att1})\; 
                $\alpha_{ij}^{\{l\}} \leftarrow$ Equation (\ref{eq:att2})\; 
                \For{$r=1,\cdots,R$}{
                    $h_{i}^{r, \{l\}} \leftarrow$ Equation (\ref{eq:att3})\;
                }
                $h_{i}^{\{l\}} \leftarrow$ Equation (\ref{eq:att4})\;
            }
            obtain node representations $\mathbf{H}^{\{L\}}$\;
        }
        $z_i^{\alpha}, z_i^{\beta} \leftarrow$ Equation (\ref{eq:proj})\;
        $\mathcal{L} \leftarrow$ Equation (\ref{eq:con1}-\ref{eq:con2})\;
        update parameters through backpropagation\;
    } 
    \Return $\mathbf{H}=\left[\mathbf{H}^{\{0\}} \| \mathbf{H}^{\{L\}}\right]$
    
\end{algorithm}

\noindent \textbf{Training Strategy.} In this article, we follow the same two-stage training mode as previous contrastive methods \cite{velickovic2019deep, zhu2020deep, you2020graph}. In the first stage, we employ the aforementioned method to obtain node representations and update parameters using $\mathcal{L}$. Due to the preservation of crucial class information in the original features \cite{chen2022towards}, in the second stage, we concatenate the encoder input $\mathbf{H}^{\{0\}}$ and the output $\mathbf{H}^{\{L\}}$: $\mathbf{H}=\left[\mathbf{H}^{\{0\}} \| \mathbf{H}^{\{L\}}\right] \in \mathbb{R}^{n\times (d_0+d_L)}$. Subsequently, we employ a simple machine learning classifier, i.e., Logistic Regression (LR) for training and testing with $\mathbf{H}$. Furthermore, due to the imbalanced distribution of classes in social bot detection, i.e., significantly fewer labeled bots compared to humans, we assign different weights to human and bot during the training process in the second stage.

\noindent \textbf{Loss Comparison.} In this paper, we delve into the different roles of cross-entropy and supervised contrastive learning losses in message passing. According to \cite{tang2022rethinking}, GNNs designed for graphs with heterophily may only act as low-pass filters due to non-uniform distributions in terms of heterophily degree. We attribute this phenomenon to the fact that cross-entropy may work well in cases with a larger number of samples, but may perform poorly in some exceptional situations, especially when training graph data has non-uniform distributions at the feature and topological levels. Relatively, contrastive loss is hardness-aware \cite{wang2021understanding}, which means that during the optimization process, it automatically focuses more on challenging negative samples. Furthermore, supervised contrastive learning, through the inclusion of labels, has the ability to bring representations of the same class closer. thereby facilitating the aggregation of more class-specific information through both homophilic and heterophilic edges. In summary, compared to cross-entropy, supervised contrastive learning is more capable of dealing with abnormal distributions in terms of heterophily.

\section{EXPERIMENTS}
In this section, we answer the following research questions.
\begin{itemize}
    \item \textbf{RQ1:} Does heterophily actually worsen the performance of earlier social bot detection techniques?
    \item \textbf{RQ2:} Does BotSCL outperform state-of-the-art methods for graph-based social bot detection?
    \item \textbf{RQ3:} How do the designed modules of BotSCL and different graph augmentation methods enhance the prediction?
    \item \textbf{RQ4:} What is the performance of BotSCL with respect to the hyperparameter?
    \item \textbf{RQ5:} Can BotSCL bring representations of test nodes from the same class closer while pushing representations of test nodes from different classes apart?
\end{itemize}

\subsection{Experiment Setup}
\subsubsection{Datasets.} Our method is graph-based and heterophily-aware, which requires relations between accounts and ground-truth labels of nodes. TwiBot-20 \cite{feng2021twibot} and TwiBot-22 \cite{feng2022twibot} containing user followers and following relations are capable of supporting our method and further experiments. Table \ref{tab:dataset} summarizes the detail of the two datasets. Compared to TwiBot-20, TwiBot-22 has a larger graph size, including more nodes and edges. We follow the same splits provided in the benchmark to ensure the results are comparable with previous works. It is worth noting that TwiBot-20 includes labels for only a small fraction of nodes, while TwiBot-22 has labels for all nodes.

\subsubsection{Baselines.} To demonstrate the effectiveness of BotSCL, we compare it with homophilic GNNs, typical heterophilic GNNs, state-of-the-art graph-based social bot detection methods, and self-supervised contrastive learning methods. 

\textbf{Homophilic GNNs:} \textbf{GCN} \cite{kipf2016semi} and \textbf{GAT} \cite{velivckovic2017graph} are typical GNNs based on the homophily assumption. Their information aggregation process can be viewed as the summation of neighbor representations.

\textbf{ Heterophilic GNNs:} \textbf{H2GCN} \cite{zhu2020beyond}, \textbf{FAGCN} \cite{bo2021beyond} and \textbf{GPRGNN} \cite{chien2020adaptive} are three models designed to mitigate the impact of heterophilic edges by employing different information aggregation strategies. Compared to homophilic GNNs, these models perform better on general graph datasets with varying degrees of heterophily.

\textbf{Graph-based Twtter Bot Detection:} 
\begin{itemize}
    \item \textbf{\cite{ali2019detect}} is the first approach to utilize graph neural networks for leveraging the graph structure information in social networks for social bot detection.
    \item \textbf{EvolveBot} \cite{yang2013empirical} extracts features such as betweenness centrality and clustering coefficient from the graph structure, and then performs classification on these features.
    \item \textbf{\cite{moghaddam2022friendship}} proposes a type of friendship preference features, which compares the features of followers with randomly collected account features from the social network.
    \item \textbf{BotRGCN} \cite{feng2021satar} use multilayer perception to mix different types of features and then input them into an R-GCN layer twice which can make full use of follower and following relations.
    \item \textbf{RGT} \cite{feng2022heterogeneity} proposes the relational graph transformer, which utilizes self-attention mechanism to aggregate information on each relation and proposes a semantic attention module to obtain information weights on different relation views. According to experiment results from \cite{feng2022twibot}, RGT achieves state-of-the-art performance among 16 different methods on TwiBot-20 relatively. 
\end{itemize} 

\begin{table}[]
\centering
\caption{The Statistic of Datasets.}
\label{tab:dataset}
\resizebox{\columnwidth}{!}{%
\begin{tabular}{@{}l|llllll@{}}
\toprule
Dataset & \#nodes & \#edges & class                & \#class                  & relation  & homo(\%) \\ \midrule
\multirow{4}{*}{TwiBot-20} & \multirow{4}{*}{229,580}   & \multirow{4}{*}{227,979}   & \multirow{2}{*}{human} & \multirow{2}{*}{5,237}   & follower & 81.44 \\
        &         &         &                      &                          & following & 33.56    \\
        &         &         & \multirow{2}{*}{bot} & \multirow{2}{*}{6,589}   & follower    & 28.99    \\
        &         &         &                      &                          & following & 75.27    \\ \midrule
\multirow{4}{*}{TwiBot-22} & \multirow{4}{*}{1,000,000} & \multirow{4}{*}{3,743,634} & \multirow{2}{*}{human} & \multirow{2}{*}{860,057} & follower & 88.05 \\
        &         &         &                      &                          & following & 96.20    \\
        &         &         & \multirow{2}{*}{bot} & \multirow{2}{*}{139,943} & follower    & 16.55    \\
        &         &         &                      &                          & following & 6.25     \\ \bottomrule
\end{tabular}%
}
\end{table}

\begin{table}[t]
\centering
\caption{Hyperparameter Setting on TwiBot-20 and TwiBot-22.}
\label{tab:hyper}
\resizebox{1\columnwidth}{!}{%
\begin{tabular}{@{}c|cc|c|cc@{}}
\toprule
Parameter         & TwiBot-20 & TwiBot-22 & Parameter & TwiBot-20 & TwiBot-22\\ \midrule
Optimizer              & AdamW     & AdamW  & Hidden     & 32        & 32    \\
LR          & 0.001     & 0.0001 & Epochs        & 200       & 50    \\
Batch             & 128       & 512    & Temperature $\tau$    & 0.07      & 0.07   \\
Layer $L$        & 2         & 2      & $\lambda^{\{1\}}$     & 1         & 1      \\
MLP dropout            & 0.5       & 0.5    & $\lambda^{\{2\}}$     & 1         & 1    \\
ATT dropout    & 0.3       & 0.3      & Balance weight        & 1:1       & 2:5  \\ \bottomrule
\end{tabular}%
}
\end{table}

\textbf{Self-supervised Contrastive Learning:} \textbf{DGI} \cite{velickovic2019deep}, \textbf{GRACE} \cite{zhu2020deep}, and \textbf{GBT} \cite{bielak2022graph} are three typical self-supervised graph contrastive learning frameworks that learn node representations in the absence of labels. \textbf{SupCon} \cite{khosla2020supervised} stands for supervised contrastive loss, and we implement it by modifying the loss function of the GRACE. 

\textbf{BotSCL:} bot detection method proposed by us. Compared to previous methods for bot detection, BotSCL not only considers both homophilic and heterophilic edges simultaneously but also introduces supervised contrastive learning instead of cross-entropy to guide message passing.

\subsubsection{Setting of hyperparameters.} The hyperparameter settings for TwiBot-20 and TwiBot-22 are shown in the Table \ref{tab:hyper}. The main difference in hyperparameter settings between the two datasets lies in the parameters related to model training. Due to the larger scale of TwiBot-22, we set it with a smaller learning rate, a larger batch size, and fewer training epochs compared to TwiBot-20. In addition, we also employ the dropout mechanism to prevent overfitting. We set the dropout rate to 0.5 for MLP and 0.3 for generated edge embeddings. Since using a small temperature coefficient in contrastive learning focuses more on challenging samples and leads to a more uniform distribution \cite{wang2021understanding}, we set the temperature coefficient to 0.07. Due to the significantly larger number of humans compared to bots in TwiBot-22, we train with a weight ratio of 2:5 for human and bot in the second stage.

\subsubsection{Implementation.} For GCN, GAT, \cite{ali2019detect}, BotRGCN, RGT and BotSCL, we use PyTorch \cite{paszke2019pytorch} and PyTorch Geometric \cite{fey2019fast} for implementation. For graph contrastive learning methods DGI, GRACE, GBT and SupCon, we implement them based on PyGCL \cite{zhu2021empirical} and only utilize the edge removing to augment the original graph. All models are running on Python3.8.12, NVIDIA Tesla A100 GPU with 40GB memory.  

\subsubsection{Evaluation Metrics.} Due to the class imbalance issue in social bot detection, we employ multiple metrics including \textbf{Accuracy}, \textbf{F1-score}, \textbf{Recall}, and \textbf{Precision} to evaluate the model performance. \textbf{Accuracy} and \textbf{F1-score} provide more insights because different methods may show significant differences in terms of Recall and Precision.

\begin{figure}[t]
    \centering
    \subfloat[Mask heterophily on TwiBot-20]{
        \centering
        \includegraphics[width=0.46\linewidth]{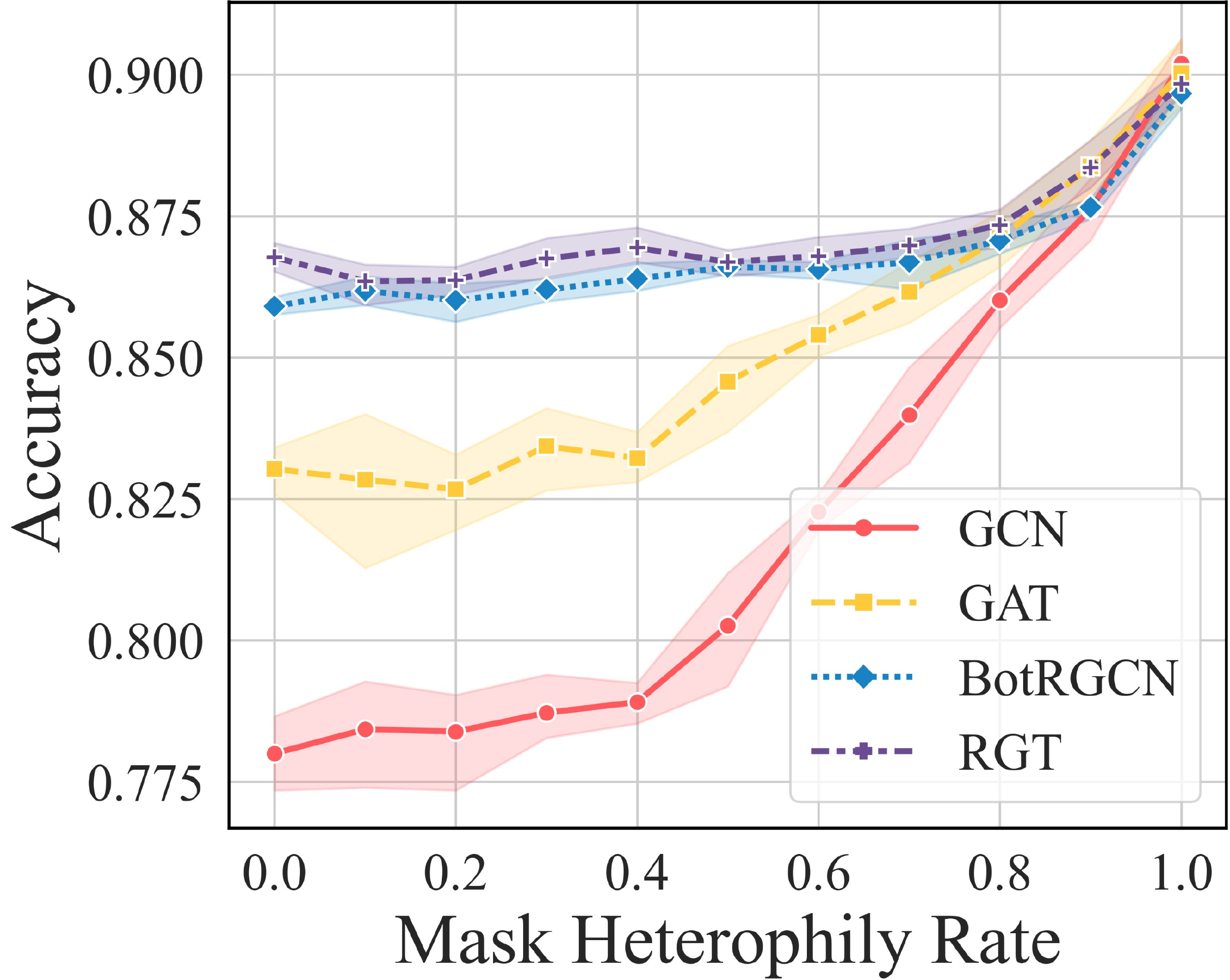}
    } 
    \hfill
    \subfloat[Mask heterophily on TwiBot-22]{
        \centering
        \includegraphics[width=0.45\linewidth]{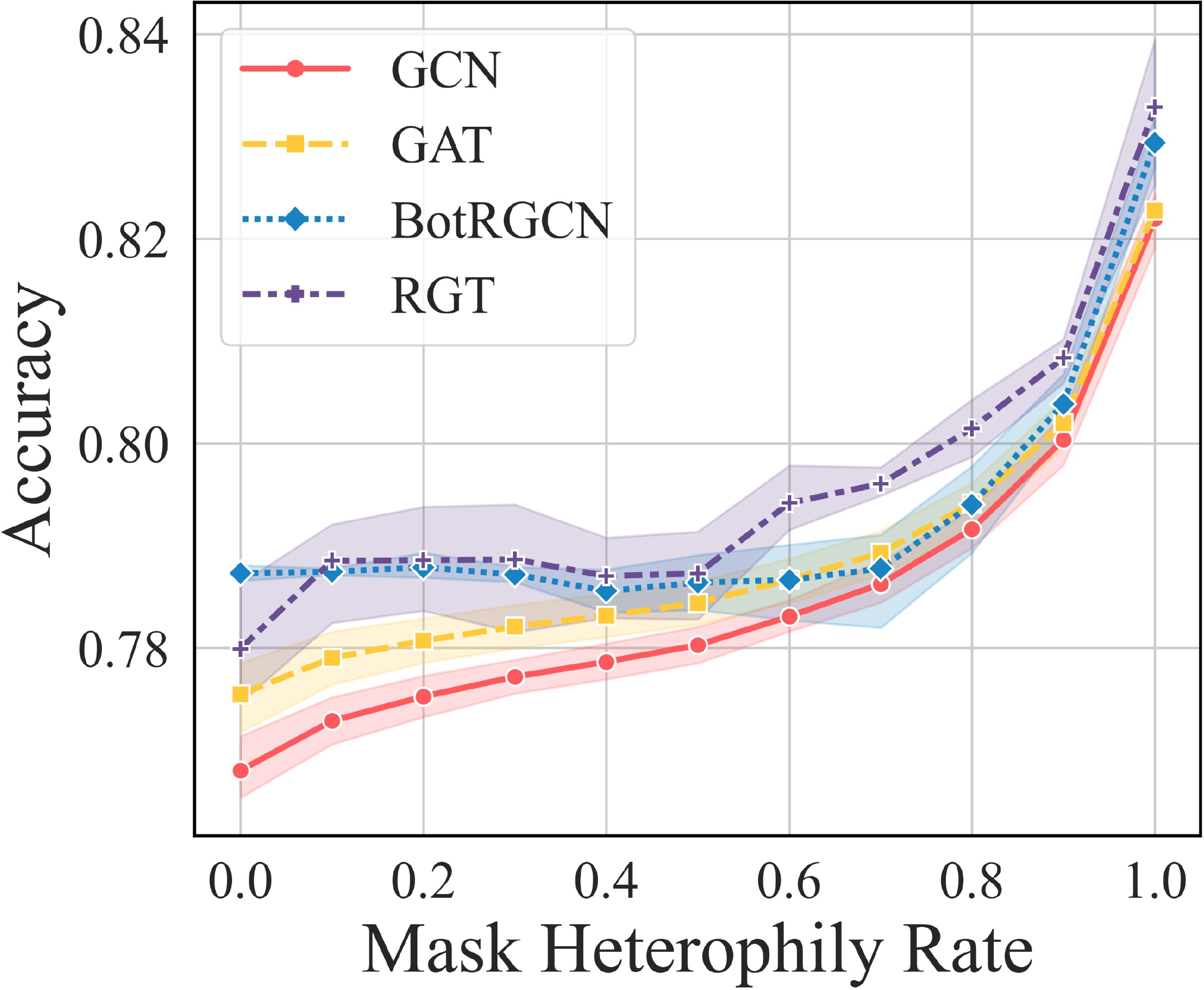}
    }
    \caption{Heterophily influence on previous graph-based methods.}
    \label{fig:rq1}
\end{figure}

\subsection{Heterophily Influence (RQ1)}
Before answering RQ1, it is necessary to analyze the degree of homophily and heterophily in two datasets. As shown in Table \ref{tab:dataset}, bots exhibit a evident heterophily tendency in the follower relation compared to human, while they display a stronger homophily tendency in the following relation in TwiBot-20. For TwiBot-22, bots show a high degree of heterophily in both relations, which can be attributed to the fact that the number of bots in the graph is significantly smaller than humans and bots are often surrounded with humans. This indicates that there is indeed an active tendency for social bots to interact more with humans. To answer RQ1, we visualize the curve of changes in accuracy as heterophilic edges are removed. We conduct experiments using four different models: GCN, GAT, BotRGCN, and RGT. As shown in Figure \ref{fig:rq1}, the accuracy of all models gradually increases as the proportion of heterophilic edges decreases with a step size of 0.1. Furthermore, we can observe that after all heterophilic edges are removed, the performance of different models is almost the same. We suspect that the core difference between these methods lies in their varying abilities to adapt to heterophily. This experiment validates that heterophilic edges are detrimental to social bot detection. Therefore, it is necessary to consider both homophilic and heterophilic edges in graph-based social bot detection.
\begin{table*}[t]
\centering
\caption{Performance Comparison on TwiBot-20 and TwiBot-22. OOM indicates Out-Of-Memory on a 40GB GPU. The best and second-best results are highlighted with \textbf{bold} and \underline{underline}.}
\label{tab:modelCompare}
\resizebox{\textwidth}{!}{%
\begin{tabular}{@{}ll|llll|llll@{}}
\toprule
\multirow{2}{*}{Methods} &
  Dataset &
  \multicolumn{4}{c|}{TwiBot-20} &
  \multicolumn{4}{c}{TwiBot-22} \\ \cmidrule(l){2-10} 
 &
  Metrics &
  \multicolumn{1}{c}{Accuracy} &
  \multicolumn{1}{c}{F1-score} &
  \multicolumn{1}{c}{Recall} &
  \multicolumn{1}{c|}{Precision} &
  \multicolumn{1}{c}{Accuracy} &
  \multicolumn{1}{c}{F1-score} &
  \multicolumn{1}{c}{Recall} &
  \multicolumn{1}{c}{Precision} \\ \midrule
\multirow{2}{*}{Homophilic} &
  GCN &
  77.53$\pm$1.73 &
  80.86$\pm$0.86 &
  87.62$\pm$3.31 &
  75.23$\pm$3.08 &
  78.39$\pm$0.09 &
  54.96$\pm$0.91 &
  44.80$\pm$1.71 &
  71.19$\pm$1.28 \\
 &
  GAT &
  83.27$\pm$0.56 &
  85.25$\pm$0.38 &
  89.53$\pm$0.87 &
  81.39$\pm$1.18 &
  79.48$\pm$0.09 &
  55.86$\pm$1.01 &
  44.12$\pm$1.65 &
  \textbf{76.23$\pm$1.39} \\ \midrule
\multirow{3}{*}{Heterophilic} &
  H2GCN &
  85.84$\pm$0.34 &
  87.57$\pm$0.15 &
  92.19$\pm$1.56 &
  83.44$\pm$1.32 &
  \multicolumn{1}{c}{OOM} &
  \multicolumn{1}{c}{OOM} &
  \multicolumn{1}{c}{OOM} &
  \multicolumn{1}{c}{OOM} \\
 &
  FAGCN &
  85.43$\pm$0.40 &
  87.36$\pm$0.32 &
  93.00$\pm$0.73 &
  82.39$\pm$0.70 &
  78.33$\pm$0.33 &
  52.37$\pm$2.25 &
  41.15$\pm$3.55 &
  72.68$\pm$2.75 \\
 &
  GPRGNN &
  86.05$\pm$0.34 &
  87.50$\pm$0.30 &
  90.25$\pm$0.29 &
  84.92$\pm$0.41 &
  78.34$\pm$0.09 &
  55.07$\pm$1.16 &
  45.15$\pm$2.21 &
  70.84$\pm$1.89 \\ \midrule
\multirow{5}{*}{Social Bot Detection} &
  \cite{ali2019detect} &
  59.88$\pm$0.59 &
  72.07$\pm$0.48 &
  \textbf{95.69$\pm$1.93} &
  57.81$\pm$0.43 &
  47.72$\pm$8.71 &
  38.10$\pm$5.93 &
  {\ul 56.75$\pm$17.69} &
  29.99$\pm$3.08 \\
 &
  EvolveBot &
  65.83$\pm$0.64 &
  69.75$\pm$0.51 &
  72.81$\pm$0.42 &
  66.93$\pm$0.61 &
  71.09$\pm$0.04 &
  14.09$\pm$0.09 &
  8.04$\pm$0.06 &
  56.38$\pm$0.41 \\
 &
  \cite{moghaddam2022friendship} &
  74.05$\pm$0.80 &
  77.87$\pm$0.71 &
  84.38$\pm$1.03 &
  72.29$\pm$0.67 &
  73.78$\pm$0.01 &
  32.07$\pm$0.03 &
  21.02$\pm$0.07 &
  67.61$\pm$0.10 \\
 &
  BotRGCN &
  85.75$\pm$0.69 &
  87.25$\pm$0.74 &
  90.19$\pm$1.72 &
  84.52$\pm$0.54 &
  79.66$\pm$0.14 &
  {\ul 57.50$\pm$1.42} &
  46.80$\pm$2.76 &
  74.81$\pm$2.22 \\
 &
  RGT &
  {\ul 86.57$\pm$0.42} &
  {\ul 88.01$\pm$0.42} &
  91.06$\pm$0.80 &
  \textbf{85.15$\pm$0.28} &
  76.47$\pm$0.45 &
  42.94$\pm$0.49 &
  30.01$\pm$0.17 &
  {\ul 75.03$\pm$0.85} \\ \midrule
\multirow{4}{*}{Contrative Learning} &
  DGI &
  84.93$\pm$0.31 &
  87.09$\pm$0.36 &
  {\ul 93.94$\pm$1.13} &
  81.17$\pm$0.26 &
  79.61$\pm$0.13 &
  44.06$\pm$1.52 &
  34.45$\pm$1.97 &
  61.28$\pm$0.64 \\
 &
  GRACE &
  84.74$\pm$0.88 &
  86.90$\pm$0.84 &
  93.56$\pm$1.57 &
  81.13$\pm$0.55 &
  {\ul 80.02$\pm$0.91} &
  46.17$\pm$4.48 &
  37.02$\pm$5.54 &
  62.05$\pm$2.84 \\
 &
  GBT &
  84.74$\pm$0.92 &
  86.87$\pm$0.79 &
  93.28$\pm$1.14 &
  81.29$\pm$0.92 &
  79.75$\pm$0.76 &
  47.27$\pm$3.08 &
  39.10$\pm$4.60 &
  60.87$\pm$3.93 \\
 &
  SupCon &
  86.10$\pm$0.14 &
  87.67$\pm$0.16 &
  91.37$\pm$0.65 &
  84.27$\pm$0.37 &
  80.00$\pm$0.24 &
  44.41$\pm$3.83 &
  34.58$\pm$5.26 &
  63.52$\pm$2.53 \\ \midrule
Ours &
  BotSCL &
  \textbf{87.26$\pm$0.31} &
  \textbf{88.79$\pm$0.27} &
  93.24$\pm$0.41 &
  {\ul 84.74$\pm$0.37} &
  \textbf{82.39$\pm$0.50} &
  \textbf{61.53$\pm$1.45} &
  \textbf{60.38$\pm$2.89} &
  62.82$\pm$1.46 \\ \bottomrule
\end{tabular}%
}
\end{table*}

\subsection{Performance Comparison (RQ2)}
Table \ref{tab:modelCompare} summarizes the detection results of all the baseline methods and our BotSCL on TwiBot-20 and TwiBot-22. It is evident that BotSCL outperforms all the other 14 baselines in both datasets. The superior performance also indicates the significance of modeling both homophily and heterophily in finding advanced bots with the capability of actively building heterophilic edges to evade detection. 

Firstly, compared to homophilic GNNs and previous graph-based social bot detection methods which simply treat interactions in social bot detection as homophilic, BotSCL has better adaptability to distinct neighborhood distributions associated with homophily and heterophily. According to Table \ref{tab:modelCompare}, heterophilic GNNs perform better than homophilic GNNs. In addition, heterophilic GNNs achieve performance similar to previous graph-based detection methods even without the leverage of multi-relation information. Taking heterophilic edges into consideration can help uncover highly sophisticated bot accounts that are closely connected with humans.

In contrast to heterophilic GNNs, BotSCL not only considers different types of relations, but also uses supervised contrastive loss as the optimization objective, leading to improvements in both accuracy and f1-score in the two datasets. Experiments conducted by \cite{feng2022heterogeneity} have shown that information from different types of relations can enhance model classification ability. Furthermore, with supervised contrastive learning, BotSCL enforces the aggregation of homophilic neighbor information and the differentiation of heterophilic neighbor representations.

The self-supervised contrastive learning methods perform poorly on TwiBot-20 with a large number of unlabeled nodes, while performing better on TwiBot-22. We think that the reason is due to the issue of uneven training distribution, as the second training stage uses fewer training nodes on TwiBot-20. Our method is significantly superior to these self-supervised methods, indicating that supervised signals play a vital role in adapting to both homophilic and heterophilic edges in social bot detection.

\begin{table}[t]
\centering
\caption{Results in terms of Different Modules and Graph Augmentation Methods in Ablation Study.}
\label{tab:ablation}
\begin{tabular}{@{}l|cc|cc@{}}
\toprule
\multirow{2}{*}{Variants} & \multicolumn{2}{c|}{TwiBot-20}  & \multicolumn{2}{c}{TwiBot-22}   \\ \cmidrule(l){2-5} 
                          & Accuracy       & F1-score       & Accuracy       & F1-score       \\ \midrule
BotSCL                    & 87.26$\pm$0.31 & 88.79$\pm$0.27 & 82.39$\pm$0.50 & 61.53$\pm$1.45 \\
w/o Sup           & 86.45$\pm$0.21 & 88.45$\pm$0.19 & 82.00$\pm$0.61 & 56.20$\pm$2.88 \\
w/o Neg                  & 86.90$\pm$0.45 & 88.48$\pm$0.35 & 82.12$\pm$0.46 & 53.43$\pm$1.16 \\
CE                        & 84.50$\pm$0.53 & 87.19$\pm$0.36 & 80.13$\pm$0.86 & 46.59$\pm$0.89 \\ \midrule
CND                       & 87.13$\pm$0.17 & 88.70$\pm$0.20 & 81.90$\pm$0.26 & 58.78$\pm$0.84 \\
EA                        & 86.96$\pm$0.14 & 88.55$\pm$0.11 & 82.02$\pm$0.64 & 58.93$\pm$2.17 \\
ER                        & 87.11$\pm$0.15 & 88.67$\pm$0.14 & 82.17$\pm$0.37 & 61.46$\pm$1.46 \\
FM                        & 86.81$\pm$0.21 & 88.50$\pm$0.14 & 81.99$\pm$0.40 & 61.78$\pm$1.29 \\ \bottomrule
\end{tabular}%
\end{table}

\begin{figure}[t]
    \centering
    \vspace{-2em}
    \subfloat[$\lambda^{\{1\}}$ and $\lambda^{\{2\}}$ on TwiBot-20]{
        \centering
        \includegraphics[width=0.95\linewidth]{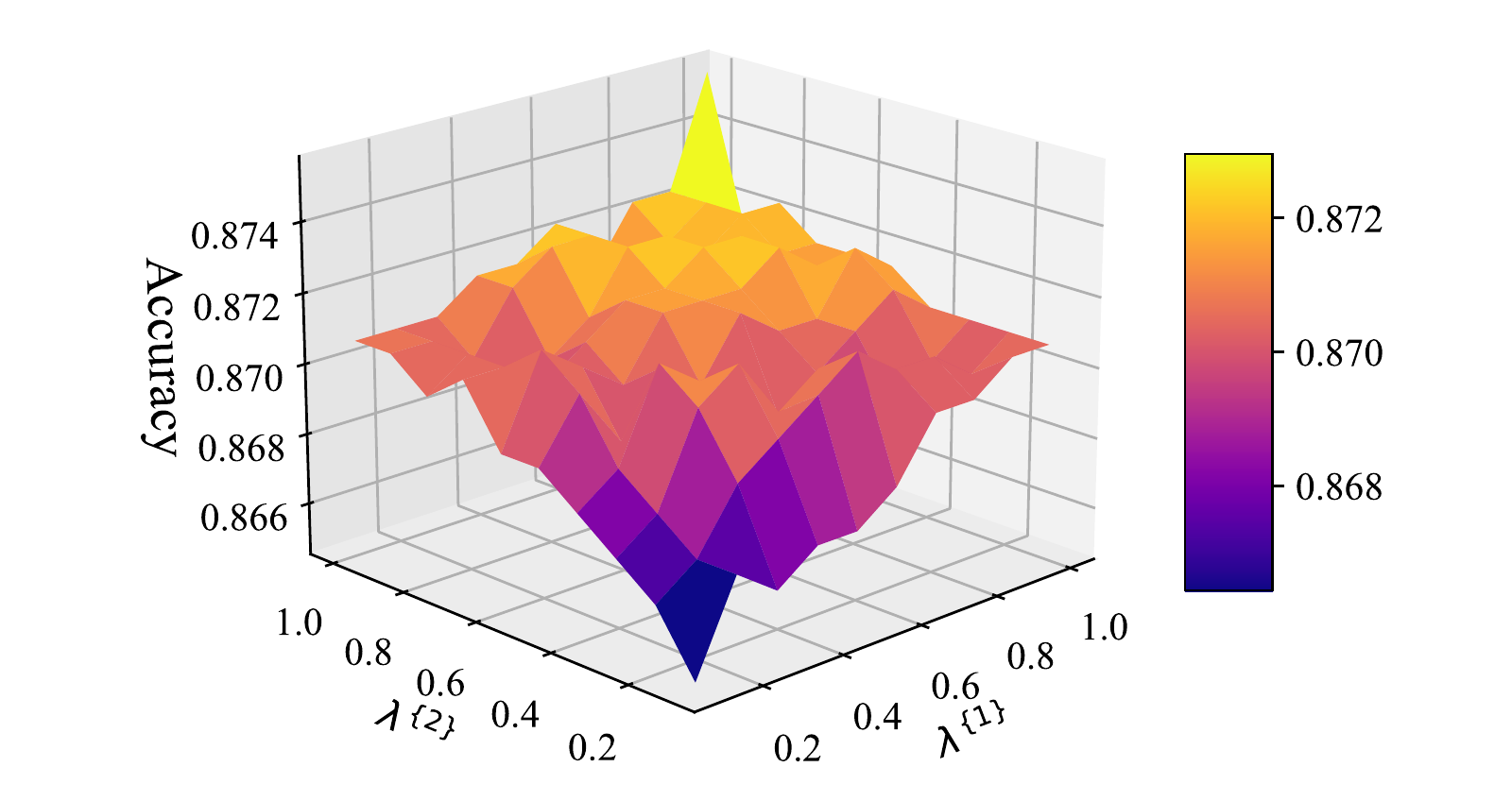}
        \label{fig:rq4a}
    }
    \hfill
    \vspace{-0.9em}
    \subfloat[$\lambda^{\{1\}}$ and $\lambda^{\{2\}}$ on TwiBot-22]{
        \centering
        \includegraphics[width=0.95\linewidth]{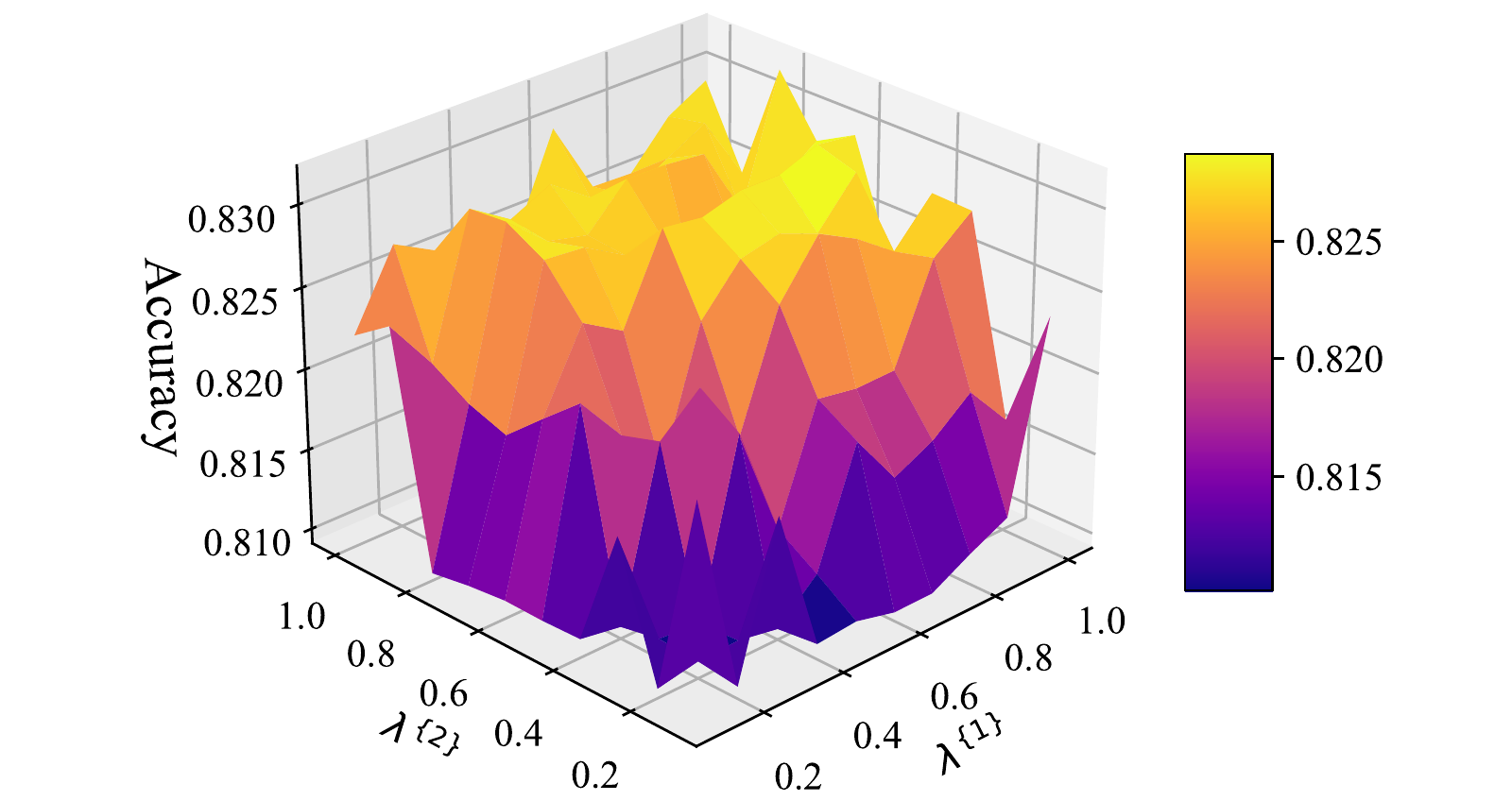}
        \label{fig:rq4b}
    }
    \caption{Sensitive Analysis of Hyperparameter $\lambda^{\{1\}}$ and $\lambda^{\{2\}}$.}
    \label{fig:rq4}
\end{figure}

\begin{figure*}[t]
    \centering
    \subfloat[GCN]{
    \centering
    \includegraphics[width=0.15\textwidth]{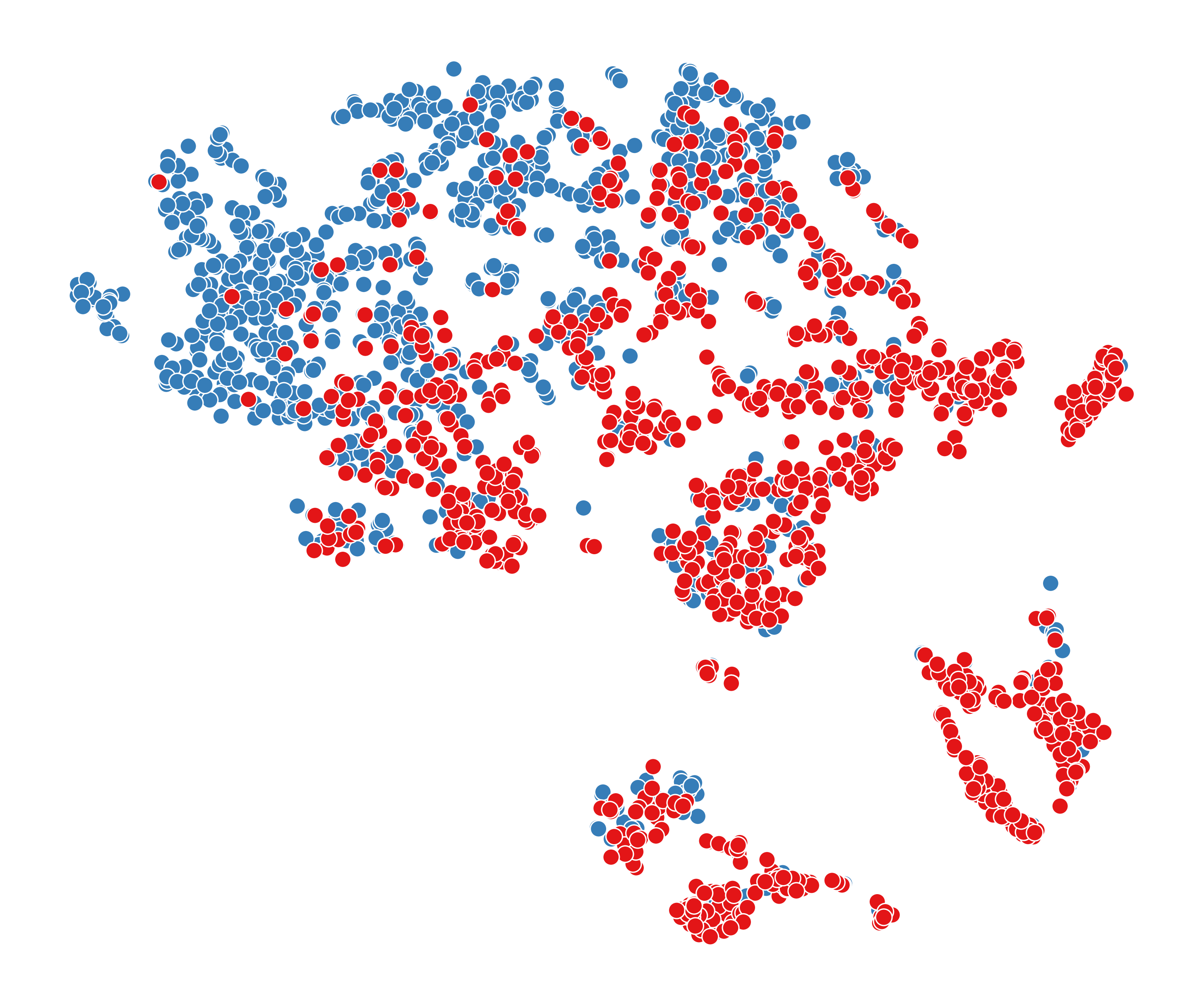}
    }
    \subfloat[FAGCN]{
    \centering
    \includegraphics[width=0.15\textwidth]{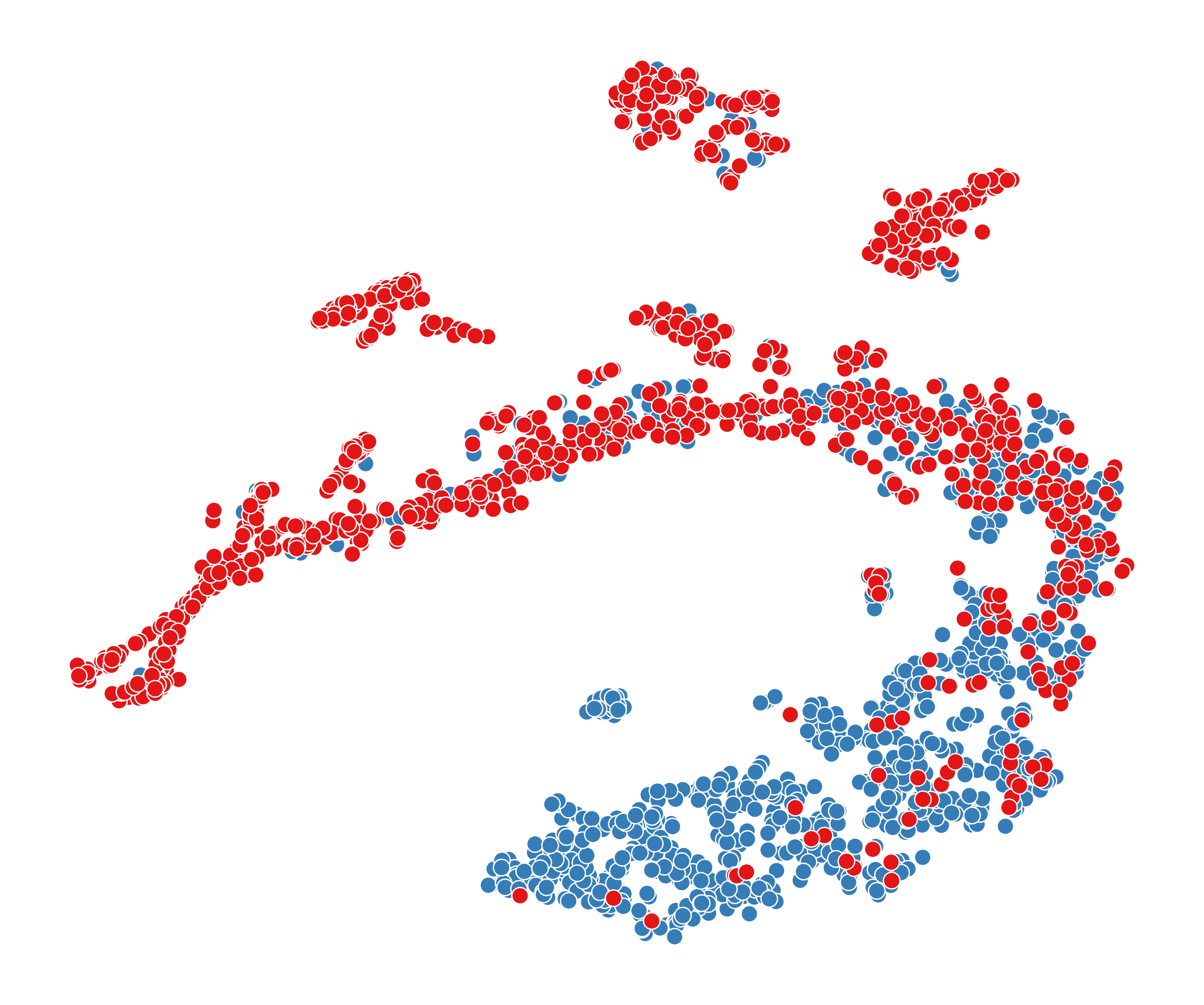}
    }
    \subfloat[BotRGCN]{
    \centering
    \includegraphics[width=0.15\textwidth]{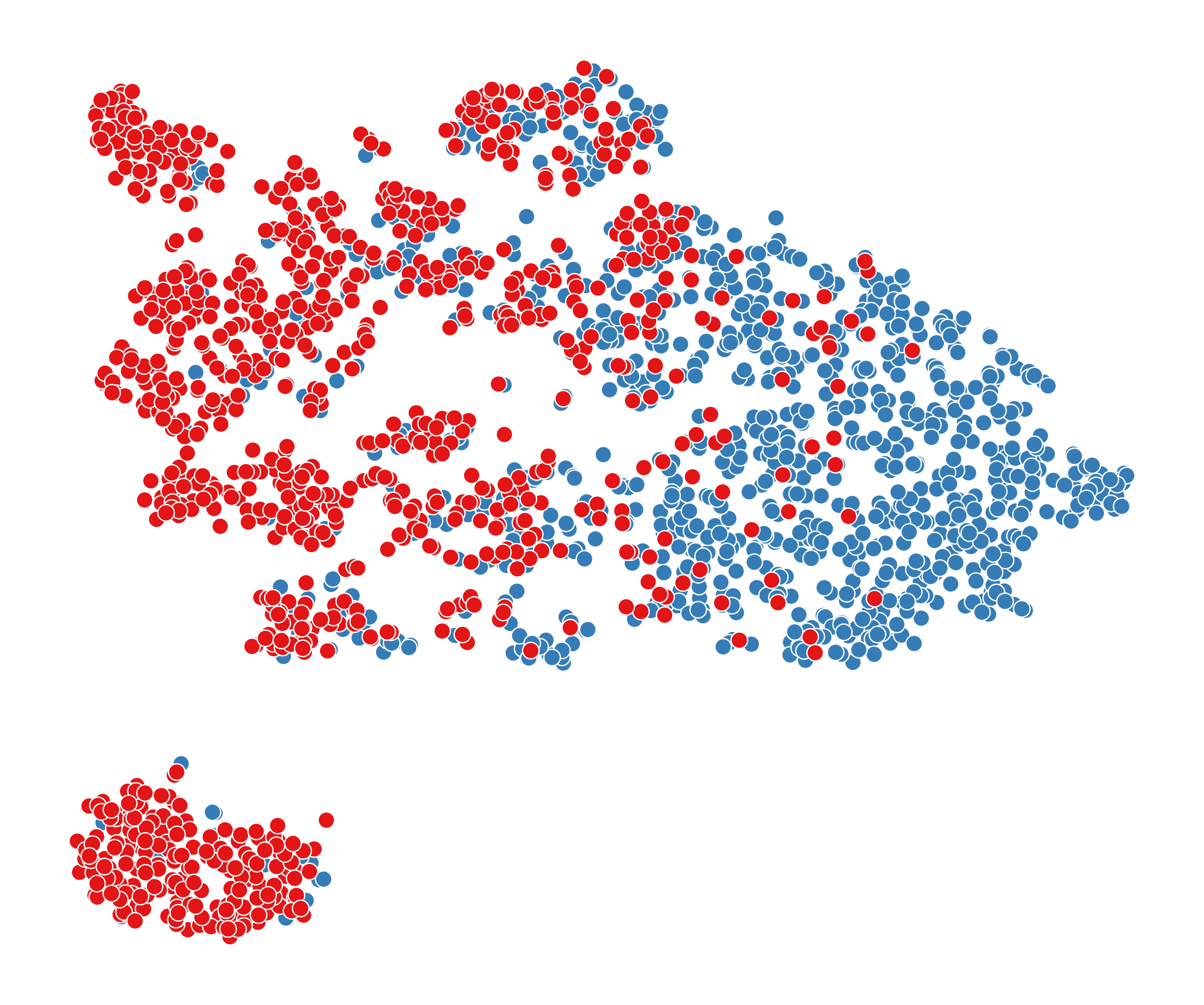}
    }
    \subfloat[RGT]{
    \centering
    \includegraphics[width=0.15\textwidth]{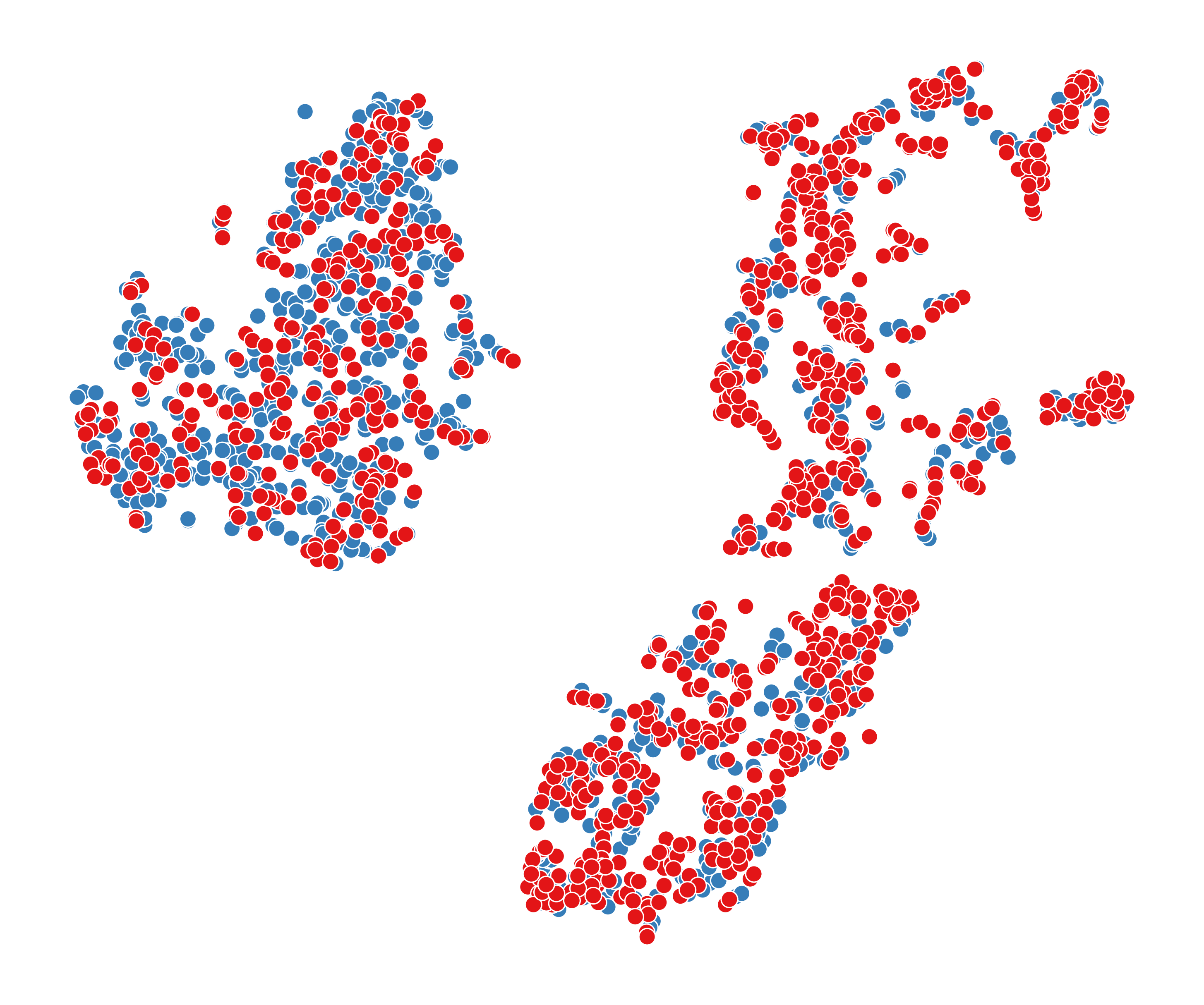}
    }
    \subfloat[DGI]{
    \centering
    \includegraphics[width=0.15\textwidth]{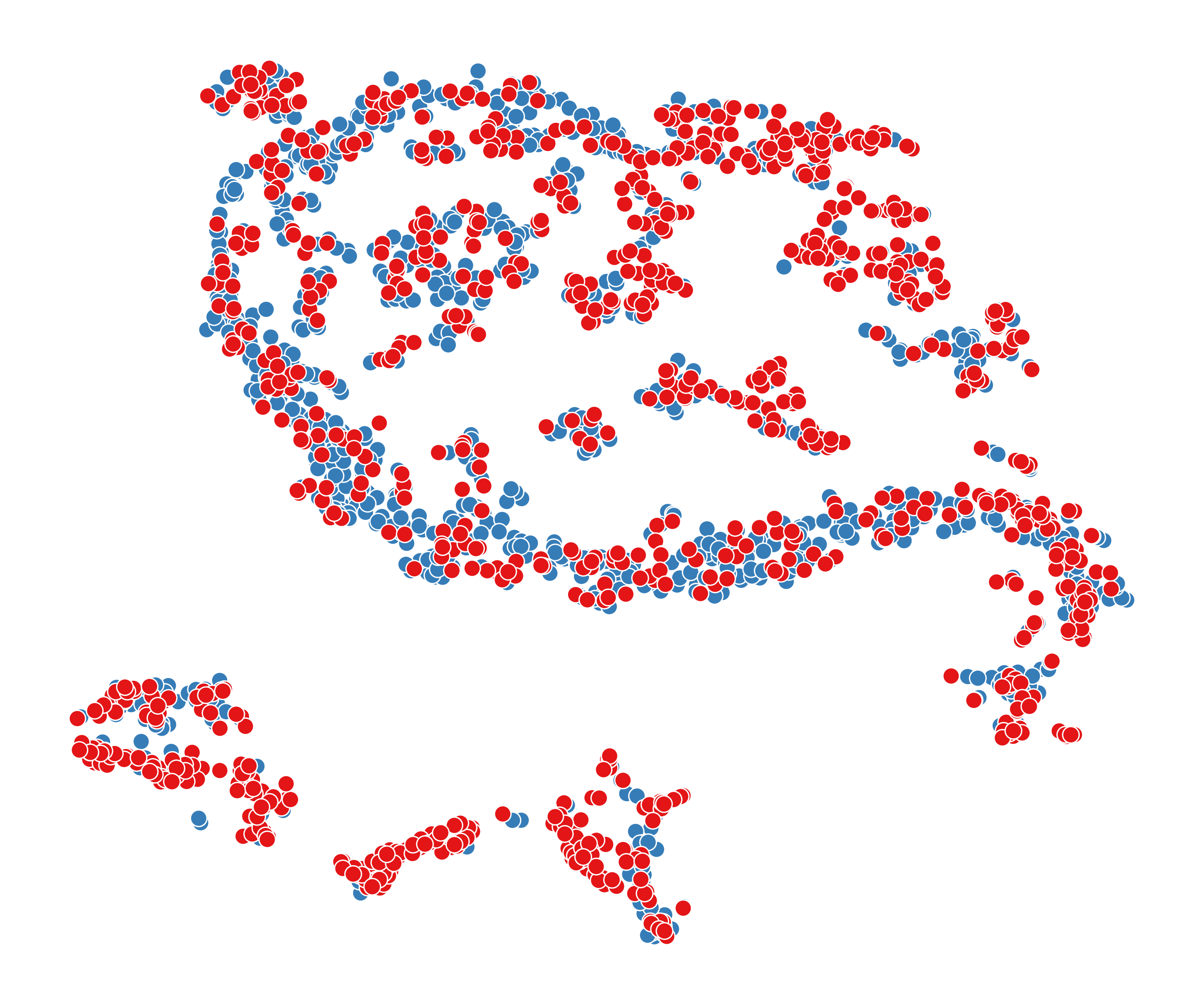}
    }
    \subfloat[BotSCL]{
    \centering
    \includegraphics[width=0.15\textwidth]{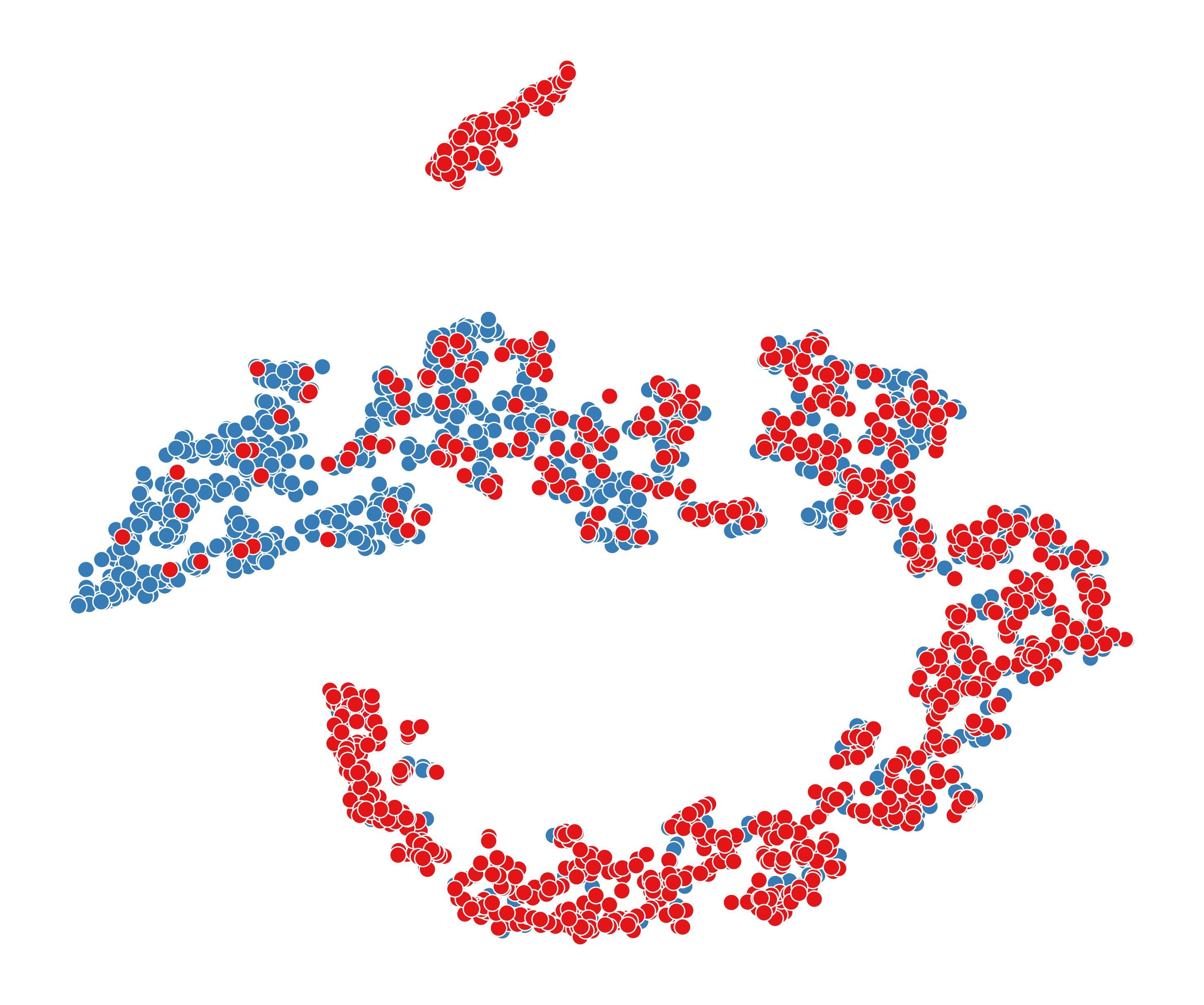}
    }
    \centering
    \caption{User Representations Visualization. \textcolor[HTML]{E31517}{Red} represents bots, while \textcolor[HTML]{367DB8}{blue} represents humans.}
    \label{fig:visualization}
    
\end{figure*}

\subsection{Ablation Study (RQ3)}
To answer RQ3, we conduct an ablation study to investigate the effects of different modules and graph augmentation methods on social bot detection. Specifically, we build three ablation models: w/o Sup, w/o Neg, and CE, developed through self-supervised contrastive learning, removal of negative information aggregation, and cross-entropy loss. We also evaluate four graph augmentation methods: CNS, Edge Adding (EA), ER, and Feature Masking (FM). The results on TwiBot-20 and TwiBot-22 are shown in Table \ref{tab:ablation}.

Compared to BotSCL, w/o Sup shows significant performance degradation, indicating that self-supervised methods struggle to aggregate similar or dissimilar information during message passing without labeled signals. The w/o Neg variant, using the softmax function for attention coefficients, aggregates similar information from neighbors but still performs well due to supervised contrastive learning, which limits aggregation from heterophilic neighbors. The CE variant suffers a notable performance drop, as cross-entropy loss focuses on classification, ignoring exceptional cases.

Additionally, different graph augmentation methods appear to have little impact on model performance, except for CNS and ER which are employed by BotSCL and show better performance. Under supervised conditions, CNS plays a crucial role in generating category-invariant representations. Among these different graph augmentation methods, CNS and ER, which BotSCL adopts, demonstrate better performance comparatively. FM and EA may potentially disrupt node feature information to some extent or introduce new noise.

\subsection{Sensitive Analysis (RQ4)}
To answer RQ4, we evaluate the performance of BotSCL with regards to hyperparameters $\lambda^{\{1\}}$ and $\lambda^{\{2\}}$, as we employ two layers of information aggregation. We keep the other parameters in the model unchanged and control the two parameters ranging from 0.1 to 1.0 with a step size of 0.1. The experimental results on TwiBot-20 and TwiBot-22 are shown in Figure \ref{fig:rq4a} and Figure \ref{fig:rq4b}, respectively.

From Figure \ref{fig:rq4},  we can clearly observe that as the hyperparameters $\lambda^{\{1\}}$ and $\lambda^{\{2\}}$ increase, the accuracy of the model on both TwiBot-20 and TwiBot-22 gradually improves.  Especially when both hyperparameters are set to 1, the model exhibits the strongest classification ability. This implies that in social bot detection, preserving the information of individual nodes is crucial for classification. Furthermore, we have found that BotSCL achieves similar performance when hyperparameters $\lambda^{\{1\}}$ and $\lambda^{\{2\}}$ have symmetric values. This could be because when they are set to symmetric values, their product remains the same, and the preservation of self-feature weights is equivalent after two layers of information aggregation. From Figure \ref{fig:rq4b}, it can be observed that as the hyperparameter decreases to 0.8, there is a significant drop in accuracy on TwiBot-22. On the other hand, on TwiBot-20, this decrease is more gradual, and a steep drop is only observed when the hyperparameter decreases to nearly 0.4.  Compared to TwiBot-20, TwiBot-22 includes labels for all nodes and suffers from the class imbalance issue, where the number of instances for each class is uneven. If the weights assigned to the original information are relatively small, the representation of the ego node is susceptible to being overwhelmed by neighbor information.

Overall, regardless of the variations in hyperparameters, the accuracy changes within the range of 0.04 for TwiBot-20 and within the range of 0.01 for TwiBot-22. This indicates that BotSCL is not highly sensitive to hyperparameters $\lambda^{\{1\}}$ and $\lambda^{\{2\}}$, although the hyperparameters do have some impact on the model's performance.

\subsection{Visualization (RQ5)}
To answer RQ5, we visualize node representations obtained from six different models, including GCN, FAGCN, BotRGCN, RGT, DGI, and our method BotSCL. Due to the larger scale of TwiBot-22, it is more representative of real-world social networks. Therefore, we take it as an example for visualization. First, we obtain node representations for each model using the same implementation method as in the comparison experiments. Then, we employ the t-SNE \cite{van2008visualizing} to reduce the dimensionality of the representations to 2D space for visualization purposes. To facilitate better observation, we randomly select 1000 human nodes and 1000 bot nodes from the test set for visualization. The visualization results of these 6 models are shown in Figure \ref{fig:visualization}. 

From Figure \ref{fig:visualization}, we can see that node representations obtained by homophilic GNNs (GCN, BotRGCN, RGT) are more scattered compared to those from FAGCN. These methods based on the homophily assumption can only smooth the representations of neighboring nodes without differentiation. As a result, the generated representations are primarily composed of local information, leading to a more scattered distribution. Among homophilic GNNs, RGT produces the poorest node representations on TwiBot-22, which is contrary to its performance on TwiBot-20. This suggests that the adaptive mechanism can be greatly influenced by the training data, leading to poor classification performance.

Compared to FAGCN, the node representations generated by BotSCL exhibit stronger clustering characteristics and include fewer local clusters. Although FAGCN takes into account both homophilic and heterophilic edges, using cross-entropy directly may inevitably overlook samples from less frequent distributions. Furthermore, the lack of consideration for multi-relational information is also one of the reasons for the suboptimal performance of FAGCN.

The node representations generated by DGI exhibit poor discriminative properties, as no supervised signals are used during the training process. The issue of class collision is also evident here, where the representations of bots and humans are almost mixed. In social bot detection, the problem of class collision in self-supervised contrastive learning is particularly severe. A large number of test nodes from different classes exhibit similar representations.

\section{Conclusion}

In this paper, we first realize that social bots can evade graph-based social bot detection methods by simply actively interacting with humans. Furthermore, through analysis of two real-world datasets, we have found strong evidence of social bots exhibiting a significant heterophily tendency. To address this, we propose BotSCL, which takes both homophilic and heterophilic edges into consideration. The key of BotSCL lies in making the encoder that can freely assimilate or discriminate neighbor representations aggregate class-specific information from neighbors. Comprehensive experiments conducted on the two real-world social bot datasets demonstrate the negative impact of heterophily on social bot detection and the effectiveness of the proposed method.

\begin{acks}
This work was supported by the National Key Research and Development Program of China through the grants 2022YFB3105405 and 2021YFC3300502, NSFC through grants 62322202 and 61932002, Beijing Natural Science Foundation through grant 4222030, Guangdong Basic and Applied Basic Research Foundation through grant 2023B1515120020, Shijiazhuang Science and Technology Plan Project through grant 231130459A.
\end{acks}

\bibliographystyle{ACM-Reference-Format}
\bibliography{main}


\begin{thebibliography}{50}


\ifx \showCODEN    \undefined \def \showCODEN     #1{\unskip}     \fi
\ifx \showDOI      \undefined \def \showDOI       #1{#1}\fi
\ifx \showISBNx    \undefined \def \showISBNx     #1{\unskip}     \fi
\ifx \showISBNxiii \undefined \def \showISBNxiii  #1{\unskip}     \fi
\ifx \showISSN     \undefined \def \showISSN      #1{\unskip}     \fi
\ifx \showLCCN     \undefined \def \showLCCN      #1{\unskip}     \fi
\ifx \shownote     \undefined \def \shownote      #1{#1}          \fi
\ifx \showarticletitle \undefined \def \showarticletitle #1{#1}   \fi
\ifx \showURL      \undefined \def \showURL       {\relax}        \fi
\providecommand\bibfield[2]{#2}
\providecommand\bibinfo[2]{#2}
\providecommand\natexlab[1]{#1}
\providecommand\showeprint[2][]{arXiv:#2}

\bibitem[Abu-El-Haija et~al\mbox{.}(2019)]%
        {abu2019mixhop}
\bibfield{author}{\bibinfo{person}{Sami Abu-El-Haija}, \bibinfo{person}{Bryan
  Perozzi}, \bibinfo{person}{Amol Kapoor}, \bibinfo{person}{Nazanin
  Alipourfard}, \bibinfo{person}{Kristina Lerman}, \bibinfo{person}{Hrayr
  Harutyunyan}, \bibinfo{person}{Greg Ver~Steeg}, {and} \bibinfo{person}{Aram
  Galstyan}.} \bibinfo{year}{2019}\natexlab{}.
\newblock \showarticletitle{Mixhop: Higher-order graph convolutional
  architectures via sparsified neighborhood mixing}. In
  \bibinfo{booktitle}{\emph{ICML}}. PMLR, \bibinfo{pages}{21--29}.
\newblock


\bibitem[Ali~Alhosseini et~al\mbox{.}(2019)]%
        {ali2019detect}
\bibfield{author}{\bibinfo{person}{Seyed Ali~Alhosseini}, \bibinfo{person}{Raad
  Bin~Tareaf}, \bibinfo{person}{Pejman Najafi}, {and}
  \bibinfo{person}{Christoph Meinel}.} \bibinfo{year}{2019}\natexlab{}.
\newblock \showarticletitle{Detect me if you can: Spam bot detection using
  inductive representation learning}. In \bibinfo{booktitle}{\emph{WWW}}.
  \bibinfo{pages}{148--153}.
\newblock


\bibitem[Beskow and Carley(2019)]%
        {beskow2019its}
\bibfield{author}{\bibinfo{person}{David~M Beskow} {and}
  \bibinfo{person}{Kathleen~M Carley}.} \bibinfo{year}{2019}\natexlab{}.
\newblock \showarticletitle{Its all in a name: detecting and labeling bots by
  their name}.
\newblock \bibinfo{journal}{\emph{Computational and mathematical organization
  theory}} \bibinfo{volume}{25}, \bibinfo{number}{1} (\bibinfo{year}{2019}),
  \bibinfo{pages}{24--35}.
\newblock


\bibitem[Bielak et~al\mbox{.}(2022)]%
        {bielak2022graph}
\bibfield{author}{\bibinfo{person}{Piotr Bielak}, \bibinfo{person}{Tomasz
  Kajdanowicz}, {and} \bibinfo{person}{Nitesh~V Chawla}.}
  \bibinfo{year}{2022}\natexlab{}.
\newblock \showarticletitle{Graph Barlow Twins: A self-supervised
  representation learning framework for graphs}.
\newblock \bibinfo{journal}{\emph{Knowledge-Based Systems}}
  \bibinfo{volume}{256} (\bibinfo{year}{2022}), \bibinfo{pages}{109631}.
\newblock


\bibitem[Bo et~al\mbox{.}(2021)]%
        {bo2021beyond}
\bibfield{author}{\bibinfo{person}{Deyu Bo}, \bibinfo{person}{Xiao Wang},
  \bibinfo{person}{Chuan Shi}, {and} \bibinfo{person}{Huawei Shen}.}
  \bibinfo{year}{2021}\natexlab{}.
\newblock \showarticletitle{Beyond low-frequency information in graph
  convolutional networks}. In \bibinfo{booktitle}{\emph{AAAI}}.
  \bibinfo{pages}{3950--3957}.
\newblock


\bibitem[Chavoshi et~al\mbox{.}(2017)]%
        {chavoshi2017temporal}
\bibfield{author}{\bibinfo{person}{Nikan Chavoshi}, \bibinfo{person}{Hossein
  Hamooni}, {and} \bibinfo{person}{Abdullah Mueen}.}
  \bibinfo{year}{2017}\natexlab{}.
\newblock \showarticletitle{Temporal patterns in bot activities}. In
  \bibinfo{booktitle}{\emph{WWW}}. \bibinfo{pages}{1601--1606}.
\newblock


\bibitem[Chen et~al\mbox{.}(2022)]%
        {chen2022towards}
\bibfield{author}{\bibinfo{person}{Jingfan Chen}, \bibinfo{person}{Guanghui
  Zhu}, \bibinfo{person}{Yifan Qi}, \bibinfo{person}{Chunfeng Yuan}, {and}
  \bibinfo{person}{Yihua Huang}.} \bibinfo{year}{2022}\natexlab{}.
\newblock \showarticletitle{Towards Self-supervised Learning on Graphs with
  Heterophily}. In \bibinfo{booktitle}{\emph{CIKM}}. \bibinfo{pages}{201--211}.
\newblock


\bibitem[Chien et~al\mbox{.}(2020)]%
        {chien2020adaptive}
\bibfield{author}{\bibinfo{person}{Eli Chien}, \bibinfo{person}{Jianhao Peng},
  \bibinfo{person}{Pan Li}, {and} \bibinfo{person}{Olgica Milenkovic}.}
  \bibinfo{year}{2020}\natexlab{}.
\newblock \showarticletitle{Adaptive universal generalized pagerank graph
  neural network}.
\newblock \bibinfo{journal}{\emph{arXiv preprint arXiv:2006.07988}}
  (\bibinfo{year}{2020}).
\newblock


\bibitem[Cresci(2020)]%
        {cresci2020decade}
\bibfield{author}{\bibinfo{person}{Stefano Cresci}.}
  \bibinfo{year}{2020}\natexlab{}.
\newblock \showarticletitle{A decade of social bot detection}.
\newblock \bibinfo{journal}{\emph{Commun. ACM}} \bibinfo{volume}{63},
  \bibinfo{number}{10} (\bibinfo{year}{2020}), \bibinfo{pages}{72--83}.
\newblock


\bibitem[Deb et~al\mbox{.}(2019)]%
        {deb2019perils}
\bibfield{author}{\bibinfo{person}{Ashok Deb}, \bibinfo{person}{Luca Luceri},
  \bibinfo{person}{Adam Badaway}, {and} \bibinfo{person}{Emilio Ferrara}.}
  \bibinfo{year}{2019}\natexlab{}.
\newblock \showarticletitle{Perils and challenges of social media and election
  manipulation analysis: The 2018 us midterms}. In
  \bibinfo{booktitle}{\emph{WWW}}. \bibinfo{pages}{237--247}.
\newblock


\bibitem[des Mesnards et~al\mbox{.}(2022)]%
        {des2022detecting}
\bibfield{author}{\bibinfo{person}{Nicolas~Guenon des Mesnards},
  \bibinfo{person}{David~Scott Hunter}, \bibinfo{person}{Zakaria el Hjouji},
  {and} \bibinfo{person}{Tauhid Zaman}.} \bibinfo{year}{2022}\natexlab{}.
\newblock \showarticletitle{Detecting bots and assessing their impact in social
  networks}.
\newblock \bibinfo{journal}{\emph{Operations Research}} \bibinfo{volume}{70},
  \bibinfo{number}{1} (\bibinfo{year}{2022}), \bibinfo{pages}{1--22}.
\newblock


\bibitem[Feng et~al\mbox{.}(2022a)]%
        {feng2022heterogeneity}
\bibfield{author}{\bibinfo{person}{Shangbin Feng}, \bibinfo{person}{Zhaoxuan
  Tan}, \bibinfo{person}{Rui Li}, {and} \bibinfo{person}{Minnan Luo}.}
  \bibinfo{year}{2022}\natexlab{a}.
\newblock \showarticletitle{Heterogeneity-aware twitter bot detection with
  relational graph transformers}. In \bibinfo{booktitle}{\emph{AAAI}},
  Vol.~\bibinfo{volume}{36}. \bibinfo{pages}{3977--3985}.
\newblock


\bibitem[Feng et~al\mbox{.}(2022b)]%
        {feng2022twibot}
\bibfield{author}{\bibinfo{person}{Shangbin Feng}, \bibinfo{person}{Zhaoxuan
  Tan}, \bibinfo{person}{Herun Wan}, \bibinfo{person}{Ningnan Wang},
  \bibinfo{person}{Zilong Chen}, \bibinfo{person}{Binchi Zhang},
  \bibinfo{person}{Qinghua Zheng}, \bibinfo{person}{Wenqian Zhang},
  \bibinfo{person}{Zhenyu Lei}, \bibinfo{person}{Shujie Yang}, {et~al\mbox{.}}}
  \bibinfo{year}{2022}\natexlab{b}.
\newblock \showarticletitle{TwiBot-22: Towards graph-based Twitter bot
  detection}.
\newblock \bibinfo{journal}{\emph{arXiv preprint arXiv:2206.04564}}
  (\bibinfo{year}{2022}).
\newblock


\bibitem[Feng et~al\mbox{.}(2021b)]%
        {feng2021satar}
\bibfield{author}{\bibinfo{person}{Shangbin Feng}, \bibinfo{person}{Herun Wan},
  \bibinfo{person}{Ningnan Wang}, \bibinfo{person}{Jundong Li}, {and}
  \bibinfo{person}{Minnan Luo}.} \bibinfo{year}{2021}\natexlab{b}.
\newblock \showarticletitle{Satar: A self-supervised approach to twitter
  account representation learning and its application in bot detection}. In
  \bibinfo{booktitle}{\emph{CIKM}}. \bibinfo{pages}{3808--3817}.
\newblock


\bibitem[Feng et~al\mbox{.}(2021c)]%
        {feng2021twibot}
\bibfield{author}{\bibinfo{person}{Shangbin Feng}, \bibinfo{person}{Herun Wan},
  \bibinfo{person}{Ningnan Wang}, \bibinfo{person}{Jundong Li}, {and}
  \bibinfo{person}{Minnan Luo}.} \bibinfo{year}{2021}\natexlab{c}.
\newblock \showarticletitle{Twibot-20: A comprehensive twitter bot detection
  benchmark}. In \bibinfo{booktitle}{\emph{CIKM}}. \bibinfo{pages}{4485--4494}.
\newblock


\bibitem[Feng et~al\mbox{.}(2021a)]%
        {feng2021botrgcn}
\bibfield{author}{\bibinfo{person}{Shangbin Feng}, \bibinfo{person}{Herun Wan},
  \bibinfo{person}{Ningnan Wang}, {and} \bibinfo{person}{Minnan Luo}.}
  \bibinfo{year}{2021}\natexlab{a}.
\newblock \showarticletitle{BotRGCN: Twitter bot detection with relational
  graph convolutional networks}. In \bibinfo{booktitle}{\emph{SNAM}}.
  \bibinfo{pages}{236--239}.
\newblock


\bibitem[Ferrara(2017)]%
        {ferrara2017disinformation}
\bibfield{author}{\bibinfo{person}{Emilio Ferrara}.}
  \bibinfo{year}{2017}\natexlab{}.
\newblock \showarticletitle{Disinformation and social bot operations in the run
  up to the 2017 French presidential election}.
\newblock \bibinfo{journal}{\emph{arXiv preprint arXiv:1707.00086}}
  (\bibinfo{year}{2017}).
\newblock


\bibitem[Fey and Lenssen(2019)]%
        {fey2019fast}
\bibfield{author}{\bibinfo{person}{Matthias Fey} {and}
  \bibinfo{person}{Jan~Eric Lenssen}.} \bibinfo{year}{2019}\natexlab{}.
\newblock \showarticletitle{Fast graph representation learning with PyTorch
  Geometric}.
\newblock \bibinfo{journal}{\emph{arXiv preprint arXiv:1903.02428}}
  (\bibinfo{year}{2019}).
\newblock


\bibitem[Hamdi(2022)]%
        {hamdi2022mining}
\bibfield{author}{\bibinfo{person}{Sami~Abdullah Hamdi}.}
  \bibinfo{year}{2022}\natexlab{}.
\newblock \showarticletitle{Mining ideological discourse on Twitter: The case
  of extremism in Arabic}.
\newblock \bibinfo{journal}{\emph{Discourse \& Communication}}
  \bibinfo{volume}{16}, \bibinfo{number}{1} (\bibinfo{year}{2022}),
  \bibinfo{pages}{76--92}.
\newblock


\bibitem[Khosla et~al\mbox{.}(2020)]%
        {khosla2020supervised}
\bibfield{author}{\bibinfo{person}{Prannay Khosla}, \bibinfo{person}{Piotr
  Teterwak}, \bibinfo{person}{Chen Wang}, \bibinfo{person}{Aaron Sarna},
  \bibinfo{person}{Yonglong Tian}, \bibinfo{person}{Phillip Isola},
  \bibinfo{person}{Aaron Maschinot}, \bibinfo{person}{Ce Liu}, {and}
  \bibinfo{person}{Dilip Krishnan}.} \bibinfo{year}{2020}\natexlab{}.
\newblock \showarticletitle{Supervised contrastive learning}.
\newblock \bibinfo{journal}{\emph{Advances in neural information processing
  systems}}  \bibinfo{volume}{33} (\bibinfo{year}{2020}),
  \bibinfo{pages}{18661--18673}.
\newblock


\bibitem[Kipf and Welling(2016)]%
        {kipf2016semi}
\bibfield{author}{\bibinfo{person}{Thomas~N Kipf} {and} \bibinfo{person}{Max
  Welling}.} \bibinfo{year}{2016}\natexlab{}.
\newblock \showarticletitle{Semi-supervised classification with graph
  convolutional networks}.
\newblock \bibinfo{journal}{\emph{arXiv preprint arXiv:1609.02907}}
  (\bibinfo{year}{2016}).
\newblock


\bibitem[Kudugunta and Ferrara(2018)]%
        {kudugunta2018deep}
\bibfield{author}{\bibinfo{person}{Sneha Kudugunta} {and}
  \bibinfo{person}{Emilio Ferrara}.} \bibinfo{year}{2018}\natexlab{}.
\newblock \showarticletitle{Deep neural networks for bot detection}.
\newblock \bibinfo{journal}{\emph{Information Sciences}}  \bibinfo{volume}{467}
  (\bibinfo{year}{2018}), \bibinfo{pages}{312--322}.
\newblock


\bibitem[Le et~al\mbox{.}(2022)]%
        {le2022socialbots}
\bibfield{author}{\bibinfo{person}{Thai Le}, \bibinfo{person}{Long Tran-Thanh},
  {and} \bibinfo{person}{Dongwon Lee}.} \bibinfo{year}{2022}\natexlab{}.
\newblock \showarticletitle{Socialbots on Fire: Modeling Adversarial Behaviors
  of Socialbots via Multi-Agent Hierarchical Reinforcement Learning}. In
  \bibinfo{booktitle}{\emph{ACM Web Conference 2022}}.
  \bibinfo{pages}{545--554}.
\newblock


\bibitem[Liu et~al\mbox{.}(2022)]%
        {liu2022beyond}
\bibfield{author}{\bibinfo{person}{Yixin Liu}, \bibinfo{person}{Yizhen Zheng},
  \bibinfo{person}{Daokun Zhang}, \bibinfo{person}{Vincent Lee}, {and}
  \bibinfo{person}{Shirui Pan}.} \bibinfo{year}{2022}\natexlab{}.
\newblock \showarticletitle{Beyond Smoothing: Unsupervised Graph Representation
  Learning with Edge Heterophily Discriminating}.
\newblock \bibinfo{journal}{\emph{arXiv preprint arXiv:2211.14065}}
  (\bibinfo{year}{2022}).
\newblock


\bibitem[Luan et~al\mbox{.}(2021)]%
        {luan2021heterophily}
\bibfield{author}{\bibinfo{person}{Sitao Luan}, \bibinfo{person}{Chenqing Hua},
  \bibinfo{person}{Qincheng Lu}, \bibinfo{person}{Jiaqi Zhu},
  \bibinfo{person}{Mingde Zhao}, \bibinfo{person}{Shuyuan Zhang},
  \bibinfo{person}{Xiao-Wen Chang}, {and} \bibinfo{person}{Doina Precup}.}
  \bibinfo{year}{2021}\natexlab{}.
\newblock \showarticletitle{Is Heterophily A Real Nightmare For Graph Neural
  Networks To Do Node Classification?}
\newblock \bibinfo{journal}{\emph{arXiv preprint arXiv:2109.05641}}
  (\bibinfo{year}{2021}).
\newblock


\bibitem[Moghaddam and Abbaspour(2022)]%
        {moghaddam2022friendship}
\bibfield{author}{\bibinfo{person}{Samaneh~Hosseini Moghaddam} {and}
  \bibinfo{person}{Maghsoud Abbaspour}.} \bibinfo{year}{2022}\natexlab{}.
\newblock \showarticletitle{Friendship Preference: Scalable and Robust Category
  of Features for Social Bot Detection}.
\newblock \bibinfo{journal}{\emph{IEEE Transactions on Dependable and Secure
  Computing}} (\bibinfo{year}{2022}).
\newblock


\bibitem[Paszke et~al\mbox{.}(2019)]%
        {paszke2019pytorch}
\bibfield{author}{\bibinfo{person}{Adam Paszke}, \bibinfo{person}{Sam Gross},
  \bibinfo{person}{Francisco Massa}, \bibinfo{person}{Adam Lerer},
  \bibinfo{person}{James Bradbury}, \bibinfo{person}{Gregory Chanan},
  \bibinfo{person}{Trevor Killeen}, \bibinfo{person}{Zeming Lin},
  \bibinfo{person}{Natalia Gimelshein}, \bibinfo{person}{Luca Antiga},
  {et~al\mbox{.}}} \bibinfo{year}{2019}\natexlab{}.
\newblock \showarticletitle{Pytorch: An imperative style, high-performance deep
  learning library}.
\newblock \bibinfo{journal}{\emph{Advances in neural information processing
  systems}}  \bibinfo{volume}{32} (\bibinfo{year}{2019}).
\newblock


\bibitem[Pei et~al\mbox{.}(2020)]%
        {pei2020geom}
\bibfield{author}{\bibinfo{person}{Hongbin Pei}, \bibinfo{person}{Bingzhe Wei},
  \bibinfo{person}{Kevin Chen-Chuan Chang}, \bibinfo{person}{Yu Lei}, {and}
  \bibinfo{person}{Bo Yang}.} \bibinfo{year}{2020}\natexlab{}.
\newblock \showarticletitle{Geom-gcn: Geometric graph convolutional networks}.
\newblock \bibinfo{journal}{\emph{arXiv preprint arXiv:2002.05287}}
  (\bibinfo{year}{2020}).
\newblock


\bibitem[Schlichtkrull et~al\mbox{.}(2018)]%
        {schlichtkrull2018modeling}
\bibfield{author}{\bibinfo{person}{Michael Schlichtkrull},
  \bibinfo{person}{Thomas~N Kipf}, \bibinfo{person}{Peter Bloem},
  \bibinfo{person}{Rianne van~den Berg}, \bibinfo{person}{Ivan Titov}, {and}
  \bibinfo{person}{Max Welling}.} \bibinfo{year}{2018}\natexlab{}.
\newblock \showarticletitle{Modeling relational data with graph convolutional
  networks}. In \bibinfo{booktitle}{\emph{European semantic web conference}}.
  Springer, \bibinfo{pages}{593--607}.
\newblock


\bibitem[Shi et~al\mbox{.}(2022)]%
        {shi2022h2}
\bibfield{author}{\bibinfo{person}{Fengzhao Shi}, \bibinfo{person}{Yanan Cao},
  \bibinfo{person}{Yanmin Shang}, \bibinfo{person}{Yuchen Zhou},
  \bibinfo{person}{Chuan Zhou}, {and} \bibinfo{person}{Jia Wu}.}
  \bibinfo{year}{2022}\natexlab{}.
\newblock \showarticletitle{H2-FDetector: A GNN-based Fraud Detector with
  Homophilic and Heterophilic Connections}. In \bibinfo{booktitle}{\emph{ACM
  Web Conference 2022}}. \bibinfo{pages}{1486--1494}.
\newblock


\bibitem[Tang et~al\mbox{.}(2022)]%
        {tang2022rethinking}
\bibfield{author}{\bibinfo{person}{Jianheng Tang}, \bibinfo{person}{Jiajin Li},
  \bibinfo{person}{Ziqi Gao}, {and} \bibinfo{person}{Jia Li}.}
  \bibinfo{year}{2022}\natexlab{}.
\newblock \showarticletitle{Rethinking Graph Neural Networks for Anomaly
  Detection}.
\newblock \bibinfo{journal}{\emph{arXiv preprint arXiv:2205.15508}}
  (\bibinfo{year}{2022}).
\newblock


\bibitem[Van~der Maaten and Hinton(2008)]%
        {van2008visualizing}
\bibfield{author}{\bibinfo{person}{Laurens Van~der Maaten} {and}
  \bibinfo{person}{Geoffrey Hinton}.} \bibinfo{year}{2008}\natexlab{}.
\newblock \showarticletitle{Visualizing data using t-SNE.}
\newblock \bibinfo{journal}{\emph{Journal of machine learning research}}
  \bibinfo{volume}{9}, \bibinfo{number}{11} (\bibinfo{year}{2008}).
\newblock


\bibitem[Vaswani et~al\mbox{.}(2017)]%
        {vaswani2017attention}
\bibfield{author}{\bibinfo{person}{Ashish Vaswani}, \bibinfo{person}{Noam
  Shazeer}, \bibinfo{person}{Niki Parmar}, \bibinfo{person}{Jakob Uszkoreit},
  \bibinfo{person}{Llion Jones}, \bibinfo{person}{Aidan~N Gomez},
  \bibinfo{person}{{\L}ukasz Kaiser}, {and} \bibinfo{person}{Illia
  Polosukhin}.} \bibinfo{year}{2017}\natexlab{}.
\newblock \showarticletitle{Attention is all you need}.
\newblock \bibinfo{journal}{\emph{Advances in neural information processing
  systems}}  \bibinfo{volume}{30} (\bibinfo{year}{2017}).
\newblock


\bibitem[Veli{\v{c}}kovi{\'c} et~al\mbox{.}(2017)]%
        {velivckovic2017graph}
\bibfield{author}{\bibinfo{person}{Petar Veli{\v{c}}kovi{\'c}},
  \bibinfo{person}{Guillem Cucurull}, \bibinfo{person}{Arantxa Casanova},
  \bibinfo{person}{Adriana Romero}, \bibinfo{person}{Pietro Lio}, {and}
  \bibinfo{person}{Yoshua Bengio}.} \bibinfo{year}{2017}\natexlab{}.
\newblock \showarticletitle{Graph attention networks}.
\newblock \bibinfo{journal}{\emph{arXiv preprint arXiv:1710.10903}}
  (\bibinfo{year}{2017}).
\newblock


\bibitem[Velickovic et~al\mbox{.}(2019)]%
        {velickovic2019deep}
\bibfield{author}{\bibinfo{person}{Petar Velickovic}, \bibinfo{person}{William
  Fedus}, \bibinfo{person}{William~L Hamilton}, \bibinfo{person}{Pietro
  Li{\`o}}, \bibinfo{person}{Yoshua Bengio}, {and} \bibinfo{person}{R~Devon
  Hjelm}.} \bibinfo{year}{2019}\natexlab{}.
\newblock \showarticletitle{Deep graph infomax.}
\newblock \bibinfo{journal}{\emph{ICLR (Poster)}} \bibinfo{volume}{2},
  \bibinfo{number}{3} (\bibinfo{year}{2019}), \bibinfo{pages}{4}.
\newblock


\bibitem[Wang and Liu(2021)]%
        {wang2021understanding}
\bibfield{author}{\bibinfo{person}{Feng Wang} {and} \bibinfo{person}{Huaping
  Liu}.} \bibinfo{year}{2021}\natexlab{}.
\newblock \showarticletitle{Understanding the behaviour of contrastive loss}.
  In \bibinfo{booktitle}{\emph{CVPR}}. \bibinfo{pages}{2495--2504}.
\newblock


\bibitem[Wang et~al\mbox{.}(2022)]%
        {wang2022powerful}
\bibfield{author}{\bibinfo{person}{Tao Wang}, \bibinfo{person}{Di Jin},
  \bibinfo{person}{Rui Wang}, \bibinfo{person}{Dongxiao He}, {and}
  \bibinfo{person}{Yuxiao Huang}.} \bibinfo{year}{2022}\natexlab{}.
\newblock \showarticletitle{Powerful graph convolutional networks with adaptive
  propagation mechanism for homophily and heterophily}. In
  \bibinfo{booktitle}{\emph{AAAI}}. \bibinfo{pages}{4210--4218}.
\newblock


\bibitem[Williams et~al\mbox{.}(2020)]%
        {williams2020homophily}
\bibfield{author}{\bibinfo{person}{Evan~M Williams}, \bibinfo{person}{Valerie
  Novak}, \bibinfo{person}{Dylan Blackwell}, \bibinfo{person}{Paul Platzman},
  \bibinfo{person}{Ian McCulloh}, {and} \bibinfo{person}{Nolan~Edward
  Phillips}.} \bibinfo{year}{2020}\natexlab{}.
\newblock \showarticletitle{Homophily and Transitivity in Bot Disinformation
  Networks}. In \bibinfo{booktitle}{\emph{SNAMS}}. IEEE, \bibinfo{pages}{1--7}.
\newblock


\bibitem[Wu et~al\mbox{.}(2021)]%
        {wu2021novel}
\bibfield{author}{\bibinfo{person}{Yuhao Wu}, \bibinfo{person}{Yuzhou Fang},
  \bibinfo{person}{Shuaikang Shang}, \bibinfo{person}{Jing Jin},
  \bibinfo{person}{Lai Wei}, {and} \bibinfo{person}{Haizhou Wang}.}
  \bibinfo{year}{2021}\natexlab{}.
\newblock \showarticletitle{A novel framework for detecting social bots with
  deep neural networks and active learning}.
\newblock \bibinfo{journal}{\emph{Knowledge-Based Systems}}
  \bibinfo{volume}{211} (\bibinfo{year}{2021}), \bibinfo{pages}{106525}.
\newblock


\bibitem[Yang et~al\mbox{.}(2013)]%
        {yang2013empirical}
\bibfield{author}{\bibinfo{person}{Chao Yang}, \bibinfo{person}{Robert
  Harkreader}, {and} \bibinfo{person}{Guofei Gu}.}
  \bibinfo{year}{2013}\natexlab{}.
\newblock \showarticletitle{Empirical evaluation and new design for fighting
  evolving twitter spammers}.
\newblock \bibinfo{journal}{\emph{IEEE Transactions on Information Forensics
  and Security}} \bibinfo{volume}{8}, \bibinfo{number}{8}
  (\bibinfo{year}{2013}), \bibinfo{pages}{1280--1293}.
\newblock


\bibitem[Yang et~al\mbox{.}(2022a)]%
        {yang2022botometer}
\bibfield{author}{\bibinfo{person}{Kai-Cheng Yang}, \bibinfo{person}{Emilio
  Ferrara}, {and} \bibinfo{person}{Filippo Menczer}.}
  \bibinfo{year}{2022}\natexlab{a}.
\newblock \showarticletitle{Botometer 101: Social bot practicum for
  computational social scientists}.
\newblock \bibinfo{journal}{\emph{Journal of Computational Social Science}}
  (\bibinfo{year}{2022}), \bibinfo{pages}{1--18}.
\newblock


\bibitem[Yang et~al\mbox{.}(2020)]%
        {yang2020scalable}
\bibfield{author}{\bibinfo{person}{Kai-Cheng Yang}, \bibinfo{person}{Onur
  Varol}, \bibinfo{person}{Pik-Mai Hui}, {and} \bibinfo{person}{Filippo
  Menczer}.} \bibinfo{year}{2020}\natexlab{}.
\newblock \showarticletitle{Scalable and generalizable social bot detection
  through data selection}. In \bibinfo{booktitle}{\emph{AAAI}},
  Vol.~\bibinfo{volume}{34}. \bibinfo{pages}{1096--1103}.
\newblock


\bibitem[Yang and Mirzasoleiman(2023)]%
        {yang2023contrastive}
\bibfield{author}{\bibinfo{person}{Wenhan Yang} {and} \bibinfo{person}{Baharan
  Mirzasoleiman}.} \bibinfo{year}{2023}\natexlab{}.
\newblock \showarticletitle{Contrastive Learning under Heterophily}.
\newblock \bibinfo{journal}{\emph{arXiv preprint arXiv:2303.06344}}
  (\bibinfo{year}{2023}).
\newblock


\bibitem[Yang et~al\mbox{.}(2022b)]%
        {yang2022rosgas}
\bibfield{author}{\bibinfo{person}{Yingguang Yang}, \bibinfo{person}{Renyu
  Yang}, \bibinfo{person}{Yangyang Li}, \bibinfo{person}{Kai Cui},
  \bibinfo{person}{Zhiqin Yang}, \bibinfo{person}{Yue Wang},
  \bibinfo{person}{Jie Xu}, {and} \bibinfo{person}{Haiyong Xie}.}
  \bibinfo{year}{2022}\natexlab{b}.
\newblock \showarticletitle{RoSGAS: Adaptive Social Bot Detection with
  Reinforced Self-Supervised GNN Architecture Search}.
\newblock \bibinfo{journal}{\emph{ACM Transactions on the Web}}
  (\bibinfo{year}{2022}).
\newblock


\bibitem[Yang et~al\mbox{.}(2023)]%
        {yang2023fedack}
\bibfield{author}{\bibinfo{person}{Yingguang Yang}, \bibinfo{person}{Renyu
  Yang}, \bibinfo{person}{Hao Peng}, \bibinfo{person}{Yangyang Li},
  \bibinfo{person}{Tong Li}, \bibinfo{person}{Yong Liao}, {and}
  \bibinfo{person}{Pengyuan Zhou}.} \bibinfo{year}{2023}\natexlab{}.
\newblock \showarticletitle{FedACK: Federated Adversarial Contrastive Knowledge
  Distillation for Cross-Lingual and Cross-Model Social Bot Detection}. In
  \bibinfo{booktitle}{\emph{ACM Web Conference 2023}}.
  \bibinfo{pages}{1314--1323}.
\newblock


\bibitem[You et~al\mbox{.}(2020)]%
        {you2020graph}
\bibfield{author}{\bibinfo{person}{Yuning You}, \bibinfo{person}{Tianlong
  Chen}, \bibinfo{person}{Yongduo Sui}, \bibinfo{person}{Ting Chen},
  \bibinfo{person}{Zhangyang Wang}, {and} \bibinfo{person}{Yang Shen}.}
  \bibinfo{year}{2020}\natexlab{}.
\newblock \showarticletitle{Graph contrastive learning with augmentations}.
\newblock \bibinfo{journal}{\emph{Advances in neural information processing
  systems}}  \bibinfo{volume}{33} (\bibinfo{year}{2020}),
  \bibinfo{pages}{5812--5823}.
\newblock


\bibitem[Zheng et~al\mbox{.}(2021)]%
        {zheng2021weakly}
\bibfield{author}{\bibinfo{person}{Mingkai Zheng}, \bibinfo{person}{Fei Wang},
  \bibinfo{person}{Shan You}, \bibinfo{person}{Chen Qian},
  \bibinfo{person}{Changshui Zhang}, \bibinfo{person}{Xiaogang Wang}, {and}
  \bibinfo{person}{Chang Xu}.} \bibinfo{year}{2021}\natexlab{}.
\newblock \showarticletitle{Weakly supervised contrastive learning}. In
  \bibinfo{booktitle}{\emph{Proceedings of the IEEE/CVF International
  Conference on Computer Vision}}. \bibinfo{pages}{10042--10051}.
\newblock


\bibitem[Zhu et~al\mbox{.}(2020b)]%
        {zhu2020beyond}
\bibfield{author}{\bibinfo{person}{Jiong Zhu}, \bibinfo{person}{Yujun Yan},
  \bibinfo{person}{Lingxiao Zhao}, \bibinfo{person}{Mark Heimann},
  \bibinfo{person}{Leman Akoglu}, {and} \bibinfo{person}{Danai Koutra}.}
  \bibinfo{year}{2020}\natexlab{b}.
\newblock \showarticletitle{Beyond homophily in graph neural networks: Current
  limitations and effective designs}.
\newblock \bibinfo{journal}{\emph{Advances in Neural Information Processing
  Systems}}  \bibinfo{volume}{33} (\bibinfo{year}{2020}),
  \bibinfo{pages}{7793--7804}.
\newblock


\bibitem[Zhu et~al\mbox{.}(2021)]%
        {zhu2021empirical}
\bibfield{author}{\bibinfo{person}{Yanqiao Zhu}, \bibinfo{person}{Yichen Xu},
  \bibinfo{person}{Qiang Liu}, {and} \bibinfo{person}{Shu Wu}.}
  \bibinfo{year}{2021}\natexlab{}.
\newblock \showarticletitle{An empirical study of graph contrastive learning}.
\newblock \bibinfo{journal}{\emph{arXiv preprint arXiv:2109.01116}}
  (\bibinfo{year}{2021}).
\newblock


\bibitem[Zhu et~al\mbox{.}(2020a)]%
        {zhu2020deep}
\bibfield{author}{\bibinfo{person}{Yanqiao Zhu}, \bibinfo{person}{Yichen Xu},
  \bibinfo{person}{Feng Yu}, \bibinfo{person}{Qiang Liu}, \bibinfo{person}{Shu
  Wu}, {and} \bibinfo{person}{Liang Wang}.} \bibinfo{year}{2020}\natexlab{a}.
\newblock \showarticletitle{Deep graph contrastive representation learning}.
\newblock \bibinfo{journal}{\emph{arXiv preprint arXiv:2006.04131}}
  (\bibinfo{year}{2020}).
\newblock


\end{thebibliography}

\end{document}